\documentclass[twocolumn]{aastex631}
\shorttitle{Classical Nova V1405 Cas Had $M_{\rm ejecta}$$>$$M_{\rm accreted}$}
\shortauthors{Schaefer}
\graphicspath{{./}{figures/}}
\begin{document}
\title{Classical Nova V1405 Cas Had $M_{\rm ejecta}$$>$$M_{\rm accreted}$ and so is Unlikely to be a Type Ia Supernova Progenitor}

\author[0000-0002-2659-8763]{Bradley E. Schaefer}
\affiliation{Department of Physics and Astronomy,
Louisiana State University, Baton Rouge, LA 70803, USA}

\begin{abstract}

Nova 2021 Cassiopeia (V1405 Cas) was an ordinary J(175) neon nova with the white dwarf mass  estimated to be $M_{\rm WD}$=0.60$\pm$0.10 $M_{\odot}$.  I found an orbital period of $P$=0.1884 days, and have tracked 20 times of photometric minima from 2013--2025.  I measure that $P$ {\it increased} from before to after the eruption with $P_{\rm pre}$=0.1883919$\pm$0.0000018 days and  $P_{\rm post}$=0.1884043$\pm$0.0000018 days, for $\Delta P$/$P$=66$_{-18}^{+21}$ parts-per-million.  With correction for the angular momentum loss by the binary during the eruption, I derive that the nova ejected $M_{\rm ejecta}$=7.5$\times$10$^{-4}$ $M_{\odot}$, with an extreme range of (2.9--40)$\times$10$^{-4}$ $M_{\odot}$.  The mass accreted during the previous eruption cycle comes from the trigger mass, and is $M_{\rm accreted}$=(1.6$\pm$0.4)$\times$10$^{-4}$ $M_{\odot}$.  V1405 Cas provides counterexamples against five claims about CV evolution that have dominated since the 1980s.  First, V1405 Cas has {\it positive} $\dot{P}$, and this is contrary to the Magnetic Braking Model.  Second, the $\Delta P$ is 20$\times$ {\it too small} to allow the system to fade into a hibernation state.  Third, $M_{\rm WD}$ is {\it decreasing} over time, as shown by $M_{\rm ejecta}$$>$$M_{\rm accreted}$ and by being a neon nova.  Fourth, V1405 Cas is not a Type Ia supernova progenitor, for the same reasons.  Fifth, the orbital period of V1405 Cas {\it increased} by $+$75 ppm from 2013--2025, as a counterexample to the pervasive idea that cataclysmic variables are universally declining in period from evolution.  V1405 Cas is the latest of recent measures of $\Delta P$ and $\dot{P}$ for 52 cataclysmic variables and 25 X-ray binaries that have together refuted all five claims.

\end{abstract}

\section{INTRODUCTION}

The orbital period ($P$) changes of interacting binaries with white dwarfs (WDs) are what define and drive the evolution and demographics of novae and cataclysmic variables (CVs).  The  interacting binaries originate at the time when the companion star comes into contact with its Roche lobe, and the evolution is driven by the period changing from relatively long-$P$ to short-$P$ (e.g., Rappaport, Verbunt, \& Joss 1983).  The effect and measure of CV evolution is the steady period change, $\frac{dP}{dT}$ or $\dot{P}$ in dimensionless units of days-per-day, from young to old binaries.  The reduction in period is inevitable because the binary can only {\it lose} angular momentum.  The dominant mechanism for angular momentum loss (AML) was early identified as a vague and never-observed-in-CVs mechanism called `magnetic braking' , so by dint of repetition and no known alternative, the `magnetic braking model' (MBM) has become the consensus default of our community since the early 1980s (Patterson 1984; Knigge et al. 2011).  A transient effect on top of the MBM is that the nova eruptions (and all CVs have nova eruptions sooner or later) will eject a shell of gas (with mass $M_{\rm ejecta}$), and this will always increase the orbital period.  

Until the last decade, the only measured case of the orbital period change across a nova eruption ($\Delta P$=$P_{\rm post}$-$P_{\rm pre}$) was for the ordinary nova BT Mon (Schaefer \& Patterson 1983), for which the period {\it increased} by 40 parts-per-million (ppm).  This $\Delta P$ served as a large part of the motivation for the nice model of CV evolution called `Hibernation' (Shara et al. 1986).  In this model, CVs have a relatively high accretion rate ($\dot{M}$) up until each nova eruption, whereupon the binary separates, driving the accretion rate to near zero, a state of hibernation, which lasts until the AML mechanism brings the companion star back into contact and restarts the high-accretion state.  The nature of the CV is determined by the orbital period, as the system cycles up and down in $\dot{M}$ superposed on top of the general long-$P$ to short-$P$ evolution.  With variations, this basic picture of CV evolution, defined and driven by $\dot{P}$ and $\Delta P$, has been the community consensus since the 1980s.

This consensus makes specific predictions and requirements for $\dot{P}$ and $\Delta P$ for all CVs.  These requirements and predictions are the single most fundamental mechanism for all branches and questions of CV evolution, because the period changes are what define and measure the evolution.  As such, the primary study of CVs should involve measures of $\dot{P}$ and $\Delta P$ for large numbers of CVs of all types.  Unfortunately, investigations on period changes have largely been ignored by our community.  

For measures of $\Delta P$, after the first case of BT Mon, no second measure was made until 2011 (Schaefer 2011).  A large part of the problem is that any measure of $\Delta P$ requires long data streams of archival data stretching back decades, and after the 1980s, few astronomers in the world had any real knowledge about the existence of the archival data, much less the ability or inclination to actually use it.  The rest of the problem is that all measures of $\Delta P$ also required persistent tedious photometry programs run for a hundred nights of telescope time spread over decades.  In practice, the only workers to measure $\Delta P$ have been myself and J. Patterson (Columbia), and it has taken decades from the program start in 1983 to reach fruition with many $\Delta P$ measures.  

For measures of $\dot{P}$ for CVs, many of the brightest CVs have had eclipse timings collected together to quantify period changes.  The data are always some collection of time series photometry to produce times of eclipses (or photometric minima) on many nights over many years, and the analysis is the equivalent of constructing an $O-C$ diagram with a parabolic fit providing the quantitative value for $\dot{P}$.  I recognize that some of these studies are not of high quality, while some have error bars and bumps that make for much scatter around the best fit parabola, and these can be greatly improved by adding yet more eclipse times.  Often, the inclusion of archival data has double or quadruple the duration of the $O-C$ curve, bringing the case out of the short-term transient effects up to century-long evolution effects.  Again, after the 1980s, most workers are not aware of the archival data, nor how to get and use the old photometry.  For the big picture questions of CV evolution, essentially all papers measuring $\dot{P}$ only have any discussion for just the {\it one} CV.  Further, most of the attention for the singular systems has been on important side issues, like trying to work out the AML mechanism or trying to find planets in orbit around the binary.  In practice, no paper has used any number of CVs to test the most fundamental predictions of the general picture of evolution.

Until the last decade, our community has been left with a horrifying lack of any effective testing of the requirements of the MBM, the Hibernation Model, and even whether CVs evolve from long-$P$ to short-$P$.  Relentless testing of the fundamental predictions is the bedrock of the science method.  A core set of many measures of $\dot{P}$ and $\Delta P$ for many CVs is needed to break out from the current blind presumptions that these effects are accurately given by the consensus models.

The first published measures of $\Delta P$ (after the BT Mon measure in 1983) were in 2011 for the two recurrent novae U Sco and CI Aql (Schaefer 2011).  As techniques improved and long-running observing programs ran to completion, I published $\Delta P$ measures for recurrent novae (RNe) T Pyx (Schaefer et al. 2013, Schaefer 2024), T CrB (Schaefer 2023a, 2025c), and a total of four U Sco eruptions (2022a, Schaefer \& Myers 2025).  I have published $\Delta P$ measures for ordinary nova DQ Her (Schaefer 2020a), RR Pic (2020b), V1017 Sgr (Salazar, LeBleu, Schaefer et al. 2017), HR Del (Schaefer 2020b), and QZ Aur (Schaefer et al 2019).  Schaefer (2025a) reports on $\Delta P$ for the unique helium nova V445 Pup.  This paper reports on $\Delta P$ for the ordinary nova V1405 Cas with its eruption in 2021.  To date, I have measured $\Delta P$ for a total of 15 eruptions.  Now we have a large sample of measures spanning all types of novae, and this database can be used to test the various models of CV evolution.

I have measured and collected $\dot{P}$ measures for 52 CVs of all types (Schaefer 2024).  Many of these measures are mine, from my own telescope time since the 1980s and from archival photographic photometry.  This includes $\dot{P}$ measures from before and after each of the nova eruptions as a necessary part of measuring $\Delta P$.  Eighteen $\dot{P}$ measures for SW Sex nova-like CVs were taken from the excellent paper of David Boyd, based primarily on his 934 eclipse times recorded with his private telescope over the last 17 years (Boyd 2023).  This work of Boyd has the vision, quality, and quantity that provides an exemplar for professional astronomers to strive for.  The remainder of the 52 CVs have the data taken from papers, one CV per paper, for which I often add further eclipse times to make a final overall $O-C$ fit.  This set of 52 CV $\dot{P}$ measures is what I will use to test the model predictions.  Further, Schaefer (2025a) collects 25 $\dot{P}$ measures for X-ray binaries (XRBs) of all types.  A conclusion from this is that the period changes do not depend on the nature of the compact accreting star.  So I have a total of 77 interacting binaries with measured $\dot{P}$ for use in model testing.

All of these measures of $\Delta P$ and $\dot{P}$ were made with the traditional plan of getting decades worth of eclipse timings from combinations of archival data and many nights of telescope time.  All this is labor-intensive and exacting, requiring decades of relentless work.  I have now performed the work for most all possible CVs for which this traditional plan can work, with only scant prospect of getting any more new measures from old novae.

This paper reports on the first measure of $\Delta P$ and $\dot{P}$ with a new plan.  The plan is to use various archived sky-survey data from {\it TESS}, AAVSO (American Association of Variable Star Observers), APASS (AAVSO Photometric All-Sky Survey), {\it Gaia}, Asteroid Terrestrial-Impact Last Alert System (ATLAS), and ASAS-SN.  From these light curves, I measure times of photometric minima in brightness from pre-eruption light curves of recent novae, and to use archived {\it TESS} data and AAVSO light curves to get post-eruption eclipse times.  This plan will work for recent novae that are brighter than $\sim$16th mag, and for which top quality sky-survey and AAVSO light curves are available.  The 2021 nova V1405 Cas is the first case for which this new plan can work.  The 9 `months' of {\it TESS} data 2019--2024 provide the cornerstone for both the pre-eruption and post-eruption period.  Nevertheless, timings from before 2017 are needed to get the pre-eruption orbital period (and the $\dot{P}$) with adequate accuracy.  The `amateur' contribution (from APASS and the AAVSO database) provides classic telescope light curves equal and better than the professional light curves.  The result is a well-sampled $O-C$ curve from 2013--2025, with this showing a well-defined kink in the year 2021 that provides the $\Delta P$.  This paper reports the $\Delta P$ and $\dot{P}$ values for V1405 Cas, and provides an analysis of the big picture tests for the various models of CV evolution.

\section{V1405 Cas}

The future nova was discovered and recognized and monitored as a variable star by Zbyn\u{e}k Henzl, and given the Czech Variable Star catalog\footnote{The reference for this is now the AAVSO's VSX catalog at \url{https://vsx.aavso.org/index.php?view=detail.top&oid=2216132}.  The original photometry was listed at the website of the Variable Star and Exoplanet Section of Czech Astronomical Society, see \url{https://var.astro.cz/en/}, but the photometry has been taken off the website and it no longer available.  The Czech Variable Star Catalog has only been published up to CzeV1228, in Skarka et al. 2017.} designation of CzeV3217.  I am impressed that this work picked out and highlighted the nova {\it before} the eruption.  This discovery was possible because the quiescent star is not faint.   The Czech photometry shows a rough sinewave that was interpreted as a double-minimum W UMa star with period 0.376938 days

V1405 Cas was discovered as a nova by Yuji Nakamura (Mie Japan) at magnitude 9.6 on 2021 March 18.  After a fast rise, the nova peaked from $V$ magnitude 7.5--8.0 for 225 days, with 9 superposed jitters (i.e., flares) with amplitudes 0.8--2.3 mags and durations 5--15 days.  The peak was in the first and biggest jitter, at $B$=5.6 and $V$=5.2 mag.  The eruption amplitude is 10.7 mag in the $V$-band.  From this peak, the light curve declined by 2 magnitudes in 166 days ($t_2$=166) and by 3 mags in 175 days ($t_3$=175).  After the flat jitter-filled peak of 225 days, the decline was fast and smooth, with the usual slowing decline.  Now, in late 2025, the nova has faded only to $V$=13.0.  The overall light curve is that of a typical slow nova with jitters, for a classification of J(175), pointing to a nova with a particularly small WD mass\footnote{All classical nova eruptions have their light curve {\it shapes} fitting into one of 7 light curve classes, as defined in Strope, Schaefer, \& Henden (2010).  The S-class shape is the basic fast-rise, fast-decline, and the decline slows at the transition, with such simple, smooth, and standard light curves being 38\% of all classical novae.  V1405 Cas is in the J-class, characterized by 2--10 jitters superposed on a relatively long and flat peak.  P-class has a plateau starting at the transition time, C-class has a cusp at the time of transition on top of the basic S-class shape, F-class has a long flat topped light curve, O-class has fast series of quasi-periodic oscillations between the peak and the transition, and D-class has a dust dip from dust-formation in the ejecta.  These classes are important for the particular mechanisms, and because the SPOC classes are uniquely high-velocity ejecta and high-$M_{\rm WD}$, while the JDF classes all have low-velocity ejecta and low-$M_{\rm WD}$ (Schaefer 2025d).  }.  The full $V$-band light curve (with 46,320 magnitudes) in the International Database of the AAVSO is shown in Figure 1.

\begin{figure}
	\includegraphics[width=1.01\columnwidth]{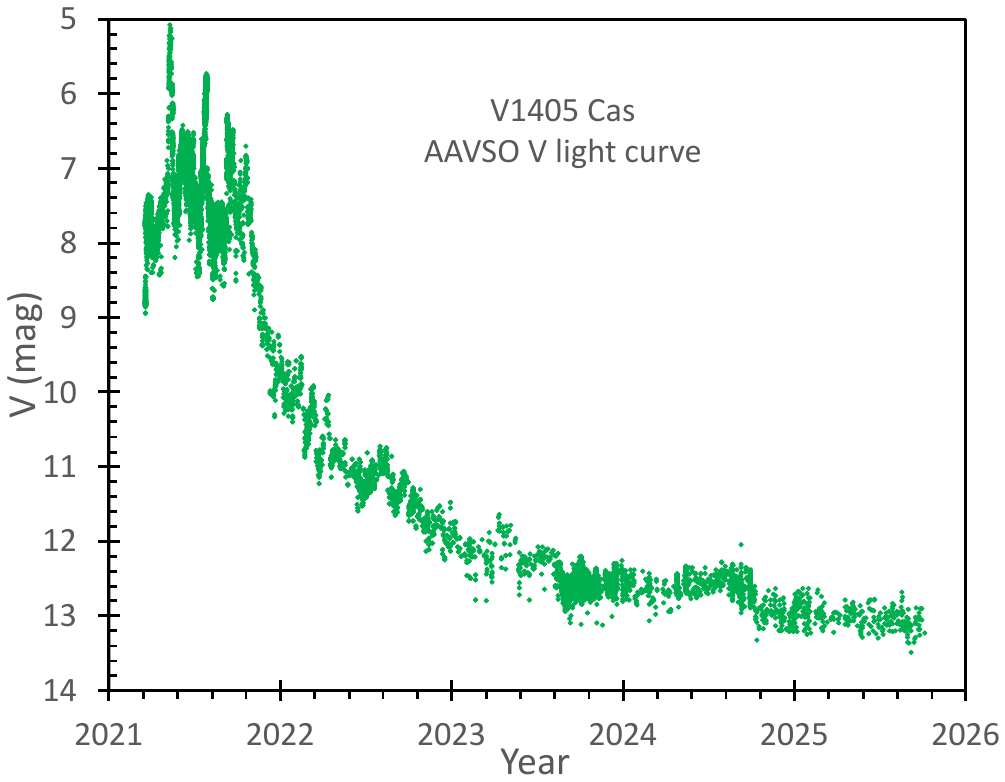}
    \caption{$V$ light curve for V1405 Cas.  This plots 46,320 magnitudes from the start of the eruption on 2021.36 up to late 2025.  These CCD magnitudes have typical error bars of 0.01 mag, with the scatter in the light curve arising from the usual flickering seen in most novae.  The orbital modulations start becoming visible only in late 2022, and these also contribute to the scatter in the plot.  This light curve class is J(175).  The J-class novae are those with a relatively flat top with multiple jitters (like flares) superposed.  V1405 Cas has jitters that peak up at $V$=5.2 mag, a relative flat-top that lasts for 225 days, and falls fast at the end of the flat-top.  The decline rate is $t_3$=175 days, which is to say that the light curve takes 175 days from the peak until the last time that it falls by 3.0 mag below the peak.  This is one of the slowest known classical novae.  The light curve class and the $t_3$ both show that the erupting white dwarf is one of the lowest known masses, which is at 0.60$\pm$0.10 $M_{\odot}$.  }
\end{figure}

During the pre-eruption quiescence, the APASS photometry gives the average pre-eruption magnitudes as $B$=15.93, $V$=15.16, and $g$=15.53.  The $V$-magnitudes vary between 15.41 and 14.73, with the variability due to the usual flickering and orbital modulation.  These magnitudes are likely an unresolved combination of V1405 Cas and two nearby background stars.  {\it Gaia} quotes $g$=15.36 for the nova and 16.52 and 16.36 for the two neighbors.  For the magnitude of the nova alone, {\it Gaia} gives $bp$=15.55, $g$=15.36, and $rp$=14.98.  With a calibration from Landolt standard stars (Carrasco et al. 2016), the difference between $g$ and $V$ magnitudes is small (with an uncertainty of roughly $\pm$0.06 mag) with $g$-$V$ near -0.04, so the average $V$ in quiescence is near 15.4 mag.  This makes the eruption amplitude into 10.2 mag.

Soon after the nova eruption, I used archival {\it TESS} light curves from three Sectors in 2019 and 2020 to recognize that the photometric periodicity was actually nearly a sinewave with half the period of the proposed W UMa star (Schaefer 2022a).  This give the period of $P$=0.1883907$\pm$0.0000048 days.  

Barrett \& Prendergast (2025), Luna, Dobrotka, \& Orio (2025), and Prendergast \& Barrett (2026) have pushed deeper into the TESS data to discover several other photometric periods at about 0.0013523 days. These periods are considered Dwarf Nova Oscillations (DNOs) by Woudt \& Warner (2002). There are also longer Quasi-Periodic Oscillations (QPOs) with the dominant period ranging between 0.0353 and 0.0728 days. The DNOs are attributed to the spin period of the WD and its beating with the orbital period. Their first and second derivatives show that the WD is slowing down rapidly as the spin period returns to its quiescent rate. The QPOs are likely attributed to instabilities in the accretion disk caused by the increased temperature and spin rate of the WD.  The WD spin, DNOs, and QPOs have no impact on the orbital period analysis that is central to this paper.

Spectroscopically, V1405 Cas started out as an ordinary nova of the He/N class, but after five weeks the spectra shifted to that typical of Fe II class, while it shifted back to the He/N class as the eruption faded, so the nova is a Hybrid (Munari, Valisa, \& Dallaporta 2021b).  Early in the eruption, reported line widths are 410 km/s for H$\beta$ (Munari, Valisa, \& Dallaporta 2021b), and 1250 km/s for the near-infrared hydrogen lines (Woodward, Banerjee, \& Evans 2021).  Late in the eruption, the H$\beta$ FWHM increased to near 2000 km/s (Di Giacomo et al. 2025), while Woodward, Starrfield, \& Page (2025) reports the Balmer lines to have FWHM of 1260 km/s.  In the X-ray spectrum, V1405 Cas remained as a bright supersoft source until at least +1546 days after the start of the eruption (Di Giacomo et al. 2025, Woodward, Starrfield, \& Page 2025).

Importantly, Munari \& Valisa (2022) found extremely strong line of [Ne III] and [Ne V] in the near-ultraviolet, so V1405 Cas is a neon nova.  Taguchi et al. (2023) found V1405 Cas to display startlingly bright Al II lines, requiring the abundance of aluminum in the ejecta to be $\sim$40$\times$ solar.  Similarly, nitrogen is $\sim$10$\times$ solar in abundance.  There is only one possible source for getting super-solar abundances of neon, aluminum, and nitrogen.  These heavy elements do not come from the normal companion star with its hydrogen rich surface, nor from nuclear burning of the accreted material (Truran \& Livio 1986).  So the only possible source is the underlying WD.  The composition of CO WDs or helium-WDs rules out any possibility of these being the source for super-solar abundances of neon, aluminum, or nitrogen (De Ger\'{o}nimo et al. 2019).  Even for the crust and mantle of ONe WDs, the Ne, Al, and N abundances are too low to provide the ejected material.  The only possible source for the Ne, Al, and N is the core of an ONe WD (De Ger\'{o}nimo et al. 2019).  The only possible way to create a stripped core with ONe composition is to start out with a freshly-made ONe WD, and strip off the outer layers with nova eruptions.  The only possibility to get the ONe WD core material into the ejecta is to have the nova eruption dredge-up the underlying mass beneath the burning accreted layer.  This means that the original ONe crust and mantle material must already have been stripped off by earlier nova eruptions.  This then requires that the current WD mass be smaller than its original mass, i.e., that $M_{\rm WD}$ is decreasing over evolutionary timescales.  This provides an independent demonstration that $M_{\rm ejecta}$$>$$M_{\rm accreted}$ for V1405 Cas.  

With the correct Bayesian priors, the {\it Gaia} parallax gives a distance of 1809$_{-179}^{+39}$ parsecs (Schaefer 2022b).  The nova is almost exactly on the Galactic plane, with a latitude of $+$0.051$\degr$.  Based on the equivalent widths of the interstellar absorption lines, $E(B-V)$=0.53 with large uncertainty (Munari, Valisa, \& Dallaporta 2021a).   The absolute magnitude at peak is $M_{\rm V,peak}$ equal to $-$7.7, which is close to the average for classical novae.  The absolute magnitude in quiescence is $M_{\rm V,q}$ equal to $+$2.5, which is relatively luminous for novae in quiescence.  The uncertainties in the absolute magnitudes are $\pm$0.3 mag, mostly from the extinction.

The mass of the WD is small (Taguchi et al. 2023, Schaefer 2025d).  V1405 Cas is one of the slowest known classical novae, so its WD mass is near the very bottom of possible values.   With the method of Shara et al. (2018), with $t_2$=166 days and $V$-band amplitude of 10.2 mag, working from Figure 1 of Schaefer (2025d), I derive $M_{\rm WD}$=0.53 $M_{\odot}$, with an uncertainty of 0.15 $M_{\odot}$.  The median Balmer line FWHM is 1260 km/s points to a small $M_{\rm WD}$.  From Equation 3 of Schaefer (2025d), the WD mass is 0.79 $M_{\odot}$, while the Figure 5 shows the error bar to be $\pm$0.20 or so.  The extreme slowness of V1405 Cas dominates over the widely varied measures of the FWHM.  The weighted average of the two mass estimates gives the WD mass to equal 0.60$\pm$0.10 $M_{\odot}$, to appropriate precision.

The mass of the companion star is best determined from its radius derived from $P$ and then converted with the CV mass-radius relation.  With Kepler's Law and $P$, for the companion star's Roche lobe being within a few percent of 0.49 $R_{\odot}$ for any plausible stellar masses.  The companion is filling its Roche lobe, so its stellar radius is close to 0.49 $R_{\odot}$.  Then, with the best mass-radius relation for CV donor stars (see Fig. 4 of Knigge et al. 2011), we get $M_{\rm comp}$=0.43$\pm$0.04 $M_{\odot}$.

I have five different means to estimate the accretion rate $\dot{M}$:  {\bf A.~}The accretion rate can be estimated from the orbital period alone, making use of the empirical $\dot{M}$ versus $P$ compilation in Gilmozzi \& Selvelli (2024).  For this, $\dot{M}$ is within a factor of 4$\times$ or so of 2$\times$10$^{-9}$ $M_{\odot}$ yr$^{-1}$.  {\bf B.~}A strict lower limit on $\dot{M}$ comes from the lack of any dwarf nova instabilities in the quiescence before the eruption, where for a period of 0.188 days requires an accretion rate $>$6$\times$10$^{-9}$ $M_{\odot}$ yr$^{-1}$ (Dubus et al. 2018).  {\bf C.~}A strict upper limit comes from the requirement that the maximum accretion is not above the limit for stable hydrogen burning on the WD surface, above which a nova eruption is impossible.  So the {\it average} accretion is far below the limit for a 0.60 $M_{\odot}$ star (see figure 7 of Shen \& Bildsten 2008), so $\dot{M}$$\ll$30$\times$10$^{-9}$ $M_{\odot}$ yr$^{-1}$.  {\bf D.~}Shara et al. (2018) have a model where the WD mass and $\dot{M}$ are uniquely determined for ordinary novae as a function of $t_2$ and the eruption amplitude.  This relation can be inverted to calculate $\dot{M}$, which turns out to be a function almost entirely of the eruption amplitude.  With a 10.2 mag amplitude, the accretion rate is 3$\times$10$^{-9}$ $M_{\odot}$ yr$^{-1}$.  For classical novae, the empirical correlation between $\dot{M}$ and eruption amplitude (see Figure 6 of Selvelli \& Gilmozzi 2019) returns the approximate rate of 10$\times$10$^{-9}$ $M_{\odot}$ yr$^{-1}$.  {\bf E.~}During quiescence, the optical light is almost entirely from the accretion disk, which is the primarily dependent on $\dot{M}$.  With $M_{\rm V,q}$=$+$2.5, the accretion disk is near the maximum for novae and CVs, pointing to an accretion rate $\sim$10$\times$10$^{-9}$ $M_{\odot}$ yr$^{-1}$ (Patterson et al. 2022).  With these five estimates, I adopt $\dot{M}$=10$\times$10$^{-9}$ $M_{\odot}$ yr$^{-1}$, with an uncertainty of roughly 50 percent.

\section{Orbital Period Changes}

The specific motivation for this paper on V1405 Cas was the realization that {\it TESS} has six Sectors of excellent post-eruption light curves, while various data sources have light curves going back to 2011.  This opens up the possibility of measuring $\Delta P$ with just archival data in hand.  So I have collected light curves from many sources, covering 2011--2025, and I have measured their times of photometric minima, covering 2013--2025:

{\it TESS} coverage is in units of $\sim$25 days of nearly gap-free light curves for a numbered Sector of time.  Nine Sectors are already downloaded, and the photometry is publicly available at the Barbara A. Mikulski Archive for Space Telescopes{\footnote{\url{https://mast.stsci.edu/portal/Mashup/Clients/Mast/Portal.html}}} (MAST).  The Sectors are 17 and 18 (October/November 2019), 24 (April 2020), 57 and 58 (October/November 2022), 77 and 78 (April/May 2024), plus 84 and 85 (October/November 2024).  The time resolution for the light curves is 1800 seconds for the three pre-eruption Sectors, and either 20 seconds or 200 seconds for the six post-eruption Sectors.  Sector 17 has 971 fluxes, while the last Sector has 10,586 fluxes.  The one-sigma measurement error (from photon noise) for each individual flux is typically $\pm$0.001 mag, so the observed light curve scatter is due to the usual flickering of a nova in quiescence.

{\it Gaia} reports 109 $G$ magnitudes (ranging from around 15.30 to 15.55) from 2014.6 to 2021.1.  (Further, 59 magnitudes from 2021.2 to 2024.0 show the fading nova.)  This light curve is publicly available on a {\it Gaia} Alerts page\footnote{\url{http://gsaweb.ast.cam.ac.uk/alerts/alert/Gaia21bpe/}}, for the target name Gaia21bpe.  The pre-eruption folded light curve is dominated by flickering, with the orbital periodicity providing only a relatively weak signal, resulting in a substantial error bar of the times of minima.

The AAVSO maintains a vast database of CCD magnitude measures in various filters, for most named variable stars\footnote{\url{https://www.aavso.org/data-download}}.  For V1405 Cas, AAVSO has 68,052 magnitudes, although many are during the eruption and before the orbital modulation starts, or in bands without substantial time series.  I selected out the $I$ and $V$ magnitudes after HJD 2460067 (when the light curve was within one mag of its quiescent level).  The $I$ band data were all taken by Richard Schmidt (observer code SREB) from 2023.86 to 2024.04, with 1874 magnitudes in total.  The $V$-band data include the $CV$ magnitudes, for which the color term is indistinguishable from zero.  I used 4867 $V$ magnitudes from 2023.33 to 2025.759, of which 72\% were from the two observers Kirill Sokolovsky (SKA) and Constantine Belyakov (BCOD), both from Russia.  The fading tail of the light curve was normalized out with a piecewise-linear trend line.  The last folded light curve (when the nova light is minimized with respect to the binary's orbital modulation) is a good sinewave with a full-amplitude of 0.20 mag.

The AAVSO has constructed its own all-sky survey in $B$, $V$, $g$, $r$, and $i$ with well-calibrated photometry of all stars down to roughly 17th mag.  The individual magnitudes from APASS are publicly available\footnote{\url{https://www.aavso.org/download-apass-data}}.  For V1405 Cas, APASS covering dates from 2011.9 to 2014.0 with 125 magnitudes in all five filters.  For light curve fitting, the colors are stable, so I have offset the magnitudes in each filter to match the average $V$ magnitude.  The best folded light curve is dominated by the ordinary flickering, but the orbital periodicity is seen as a sinewave that is significant at the 4-$\sigma$ confidence level.

The All-sky Automated Survey for Supernovae (ASAS-SN) has the light curve of V1405 Cas publicly available\footnote{\url{https://asas-sn.osu.edu/?ra=351.19887&dec=61.18744}}.  Their light curve has 157 $V$ magnitudes from 2014.9 to 2018.7, with typical photometric error bars of $\pm$0.04 mag.  The folded light curves show a sinewave (significant at the 5.6-$\sigma$ confidence level) dominated by flickering.

The Asteroid Terrestrial-Impact Last Alert System\footnote{\url{https://atlas.fallingstar.com/}} (ATLAS) has the pre-eruption light curve from 2015.5 to 2021.1 and a good long-term eruption light curve from 2021.4 to 2024.9.  The tabulated ATLAS light curve has a substantial number of far outliers, and 18\% of the magnitudes are reported as a faint mode (fainter than 15.35 mag) that does not display any periodic modulation.  After rejecting these magnitudes, the pre-eruption light curve has 1248 magnitudes, showing a periodic modulation with amplitude 0.06 mag.

I have searched widely, but can find no other light curves for which the minimum times could be measured:  The Zwicky Transient Factory has no archived source in the area of the sky around V1405 Cas.  This nova is too far north for the ``Stony Brook/SMARTS Atlas of (mostly) Southern Novae'' of F. Walter.  Pan-STARRS has too few good magnitudes to be useful.  V1405 Cas was too faint to be detected by ASAS.  The original light curve of Zbyn\u{e}k Henzl (as used to discover the photometric periodicity) has been taken off the Czech Astronomical Society website, for unknown reasons, and is not now publicly available.  V1405 Cas is outside the footprint for the Sloan Digital Sky Survey.  SuperWASP has 106 magnitudes (with error bars mostly 0.07--0.14 mag) from 2006--2008, but the periodicity is not significant.  The Harvard plates, as measured by the DASCH program, have 20 positive detections, with $\langle B \rangle$=15.2, with no prior eruptions detected and no chance of pulling out the photometric modulation.

With the good light curves, I have run Fourier transforms on each, with such confirming the 0.1884 day photometric periodicity.  The luminosity of the underlying 0.43 $M_{\odot}$ companion is greatly smaller than the disk luminosity, so any ellipsoidal modulation is not measurable.  The observed modulation can only be from the extra light from the hot spot and the irradiated side of the companion making for a beaming pattern that rotates with the binary.  So the orbital period equals the photometric period.  

\begin{figure}
	\includegraphics[width=1.01\columnwidth]{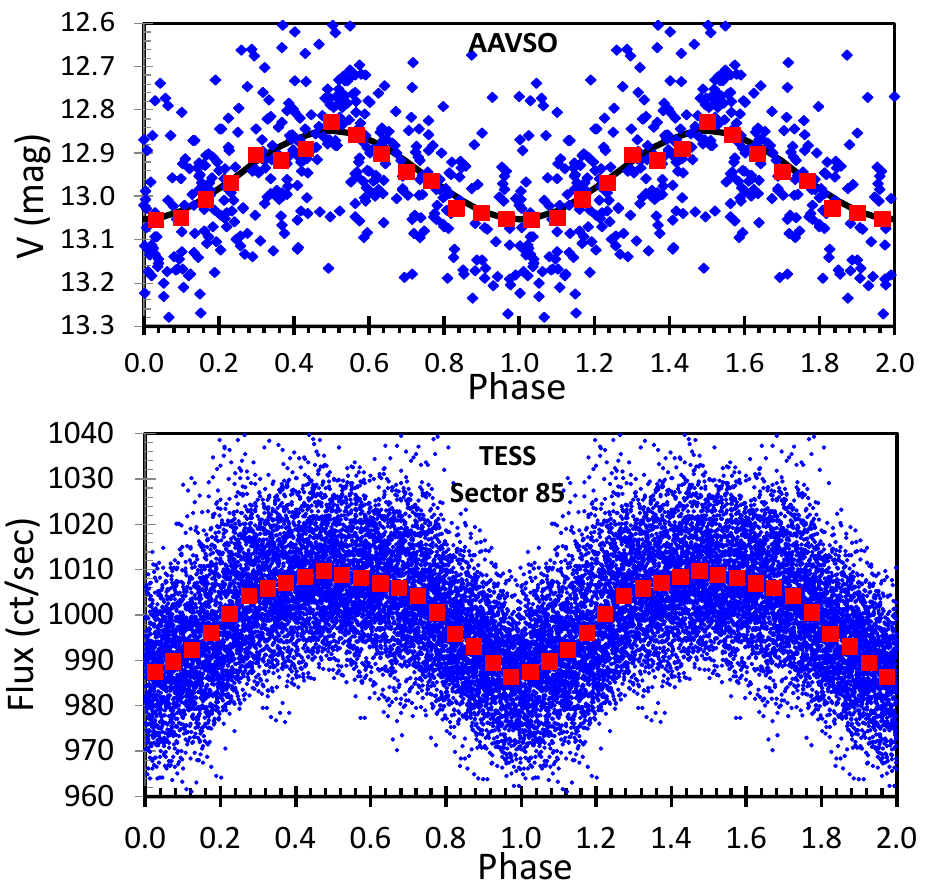}
    \caption{Folded light curves from AAVSO and {\it TESS} in late 2024.  The two panels show the folded detrended light curves for V1405 Cas, with the individual measures shown as small blue diamonds, and the phase averaged light curves shown with the larger red squares.  The upper panel is for the 412 $V$ magnitudes after 2024.79, with the phase calculated for a zero epoch of HJD 2460756.0703.  The phase averaged light curve is indistinguishable from the best fitted sinewave (shown as a black curve).  The typical CCD magnitudes have error bars of $\pm$0.01 mag, so the observed scatter is entirely from the usual flickering of a nova near quiescence.  The lower panel is for the 10586 fluxes in {\it TESS} Sector 85 (November 2024), with the phase calculated for a zero epoch of BJD 2460623.0610.  The folded light curve shape is close to a sinewave, but the minimum is somewhat narrower, while the maximum is relatively flat and broad.  The amplitude of modulation for the {\it TESS} light curves cannot be readily translated into magnitudes, because the flux measures include substantial light from background stars in the large photometry aperture.  The typical error bar is $\pm$0.9 counts/second, so the scatter is entirely from flickering in the star.  }
\end{figure}

The folded detrended $V$-band light curve from AAVSO is shown in the upper panel of Figure 2.  We see sinusoidal modulation, with a full amplitude of 0.20 mag.  The folded light curve from {\it TESS} Sector 85 is shown in the lower panel of Figure 2.  The ordinary flickering makes for a large scatter, but the phase averaged light curve shows a smooth modulation.  The light curve shape has a slightly flattened maximum and a somewhat pointy minimum.  This shape varies with the Sector.  Before the eruption, the Sectors show folded light curve indistinguishable from a sinewave.  After the eruption, the Sectors show various shapes with a somewhat flatted maximum, and some are suggestive of a weak secondary eclipse.  The full amplitude in Sector 85 is 22 in flux units, or 0.022\%.  The amplitudes in the {\it TESS} folded light curves cannot be readily converted to magnitudes, because the large pixel size (21"$\times$21") and the large photometry aperture contain a large fraction of light that is from nearby background stars.  Further, early in the tail of the nova eruption, the light from the expanding shell swamps the modulation of the binary light.

With these light curves, I have fitted templates (usually sinewaves) to the light curves.  The significant fit parameters are the period, the epoch of minimum, the amplitude, and the average magnitude.  The epoch of minimum is chosen for being a time near the middle of the observing interval (with this avoiding skewed and elongated error regions), even though this need not correspond to any specific time with an observation.  The fit is with chi-square minimization, where the best-fit is the parameter set that makes for the smallest chi-square.  The one-sigma error bars are easy to quantify as the region of parameter space over which the calculated chi-square is within 1.00 of the minimum.  For the calculation of chi-square, the $\sigma$ in the denominator should be the photometric noise and the flickering noise as added in quadrature.  In practice, the flickering noise is varied until the reduced chi-square is near unity.  The {\it TESS} data are easily divided up by Sector, to each provide one minimum time and one point in the $O-C$ curve.  The AAVSO $V$ and {\it Gaia} light curves can be broken up into multiple parts, to provide better time resolution.  The result of all this is 20 times of minimum light, $T_{\rm min}$ as a heliocentric Julian Date (HJD), including 7 pre-eruption and 10 post-eruption times, with these listed in Table 1.

\begin{table}
	\centering
	\caption{Minimum times for V1405 Cas}
	\begin{tabular}{llrr}
		\hline
		Source   &  Minimum time $T_{\rm min}$  &   $N$   &  $O-C$ \\
		   &  (HJD)  &     &  (days) \\
		\hline
APASS	&	2456369.1342	$\pm$	0.0096	&	-15802	&	0.0240	\\
ASAS-SN	&	2457717.0483	$\pm$	0.0071	&	-8647	&	0.0027	\\
{\it Gaia}	&	2457890.1906	$\pm$	0.0178	&	-7728	&	0.0139	\\
ATLAS	&	2457923.1432	$\pm$	0.0069	&	-7553	&	-0.0018	\\
ATLAS	&	2458523.1689	$\pm$	0.0051	&	-4368	&	-0.0005	\\
{\it TESS} 17	&	2458776.1768	$\pm$	0.0028	&	-3025	&	-0.0013	\\
{\it TESS} 18	&	2458804.0588	$\pm$	0.0030	&	-2877	&	-0.0012	\\
{\it TESS} 24	&	2458970.0315	$\pm$	0.0040	&	-1996	&	-0.0007	\\
{\it Gaia}	&	2459056.1174	$\pm$	0.0093	&	-1539	&	-0.0093	\\
ATLAS	&	2459103.0239	$\pm$	0.0046	&	-1290	&	-0.0121	\\
{\it TESS} 57	&	2459868.1220	$\pm$	0.0218	&	2771	&	0.0314	\\
{\it TESS} 58	&	2459897.1530	$\pm$	0.0032	&	2925	&	0.0502	\\
AAVSO $V$	&	2460216.1020	$\pm$	0.0019	&	4618	&	0.0537	\\
AAVSO $I$	&	2460287.1358	$\pm$	0.0015	&	4995	&	0.0643	\\
{\it TESS} 77	&	2460408.0914	$\pm$	0.0008	&	5637	&	0.0730	\\
{\it TESS} 78	&	2460442.1891	$\pm$	0.0008	&	5818	&	0.0720	\\
AAVSO $V$	&	2460487.0349	$\pm$	0.0028	&	6056	&	0.0808	\\
{\it TESS} 84	&	2460598.1924	$\pm$	0.0005	&	6646	&	0.0878	\\
{\it TESS} 85	&	2460623.0610	$\pm$	0.0004	&	6778	&	0.0888	\\
AAVSO $V$	&	2460756.0703	$\pm$	0.0025	&	7484	&	0.0943	\\
		\hline
	\end{tabular}		
\end{table}

These 20 observed times of minimum light ($T_{\rm min}$) can be used to construct an $O-C$ diagram.  This is a plot of the time deviations of the observed minima (`O') minus the calculated time (`C') based on some fiducial linear ephemerides.  The linear ephemerides I have adopted is an epoch close to the time of the explosion as HJD 2459346.0600 and a period of 0.1883907 days.  The ephemeris time is 
\begin{equation}
T_{\rm eph} = 2459346.0600 +N \times 0.1883907.
\end{equation}
$N$ is a cycle count since the start of the eruption.  The $O-C$ equals $T_{\rm min}$-$T_{\rm eph}$.  The full $O-C$ curve for V1405 Cas is displayed in Figure 3.  For interpretation, the {\it slope} of the $O-C$ curve is the instantaneous orbital period (as different from the ephemeris period).  A flat $O-C$ indicates a steady period at the ephemeris period, while an {\it upward} sloping straight line indicates a steady period at with a period {\it longer} than the ephemeris period, and a  {\it downward} sloping straight line indicates a steady period with a period {\it shorter} than the ephemeris period.  A steadily {\it increasing} period is depicted as a concave-{\it up} parabola, while a steadily {\it decreasing} period is a concave-{\it down} parabola.  A sudden period {\it increase} is seen as a sharp kink {\it upwards}, while a sudden period {\it decrease} is a sharp kink {\it downwards}.

\begin{figure}
	\includegraphics[width=1.01\columnwidth]{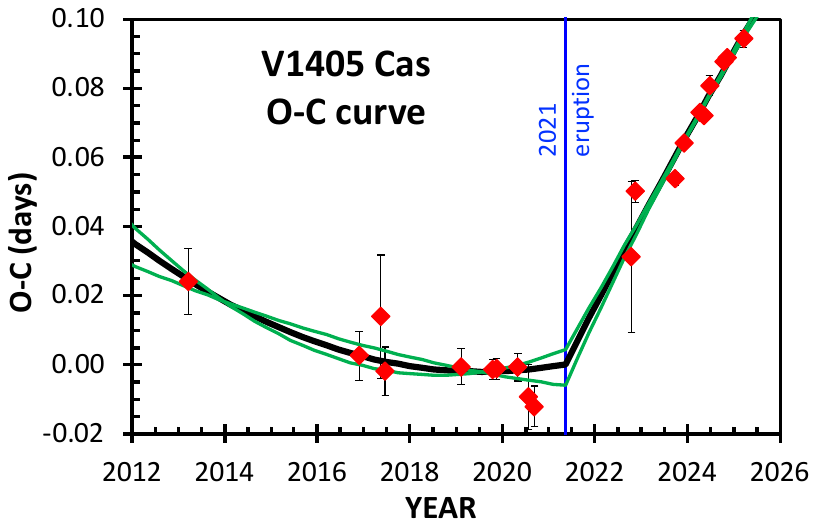}
    \caption{$O-C$ curve for V1405 Cas.  This plots 20 minimum times (red diamonds), with two points lying on top of each other in late 2024, two points overlapping in early 2024, and two points in late 2019 lying on top of each other.  The $O-C$ is calculated for an ephemeris with an epoch of 2459346.0600 and a period of 0.1883907 days.  The year of the eruption was 2021.36, as represented with the vertical blue line.  The best fitting model (Equation 2) is a broken parabola, where the curvatures show the $\dot{P}$, and the kink in 2021 shows the $\Delta P$.  The one-sigma extreme fits are shown by the two green curves.  The point of this figure is to illustrate that V1405 Cas suffered a sharp period change in 2021, and this has the orbital period {\it increase} by $+$66 parts-per-million.  }
\end{figure}

The $O-C$ curve in Figure 3 shows a fairly steady downward line before the 2021.36 eruption, and a strongly upward tilted line after the eruption, making an upward kink at the time of the eruption.  This needs to be quantified.  The way to do this is to make a chi-square fit to a model with a kink.  To be specific, the model is 
\begin{eqnarray}
T_{\rm model} = E_0 +NP_{\rm post} +0.5P\dot{P}_{\rm post}  N^2  ~~~~~~(N \ge 0), \nonumber \\
T_{\rm model} = E_0 +NP_{\rm pre} +0.5P\dot{P}_{\rm pre} N^2  ~~~~~~(N < 0).~~~
\end{eqnarray}
The nova eruption starts for $N$=0.  The epoch for the two parts of this ephemeris must be the same, $E_0$, because the companion star cannot suddenly jump forward or backward in its orbital position.  The period change will actually be spread over the duration of the eruption, but this effect makes for an unobservable negligibly-small difference in the epochs for the two parts of Equation 2, as demonstrated by the high-accuracy measures for T Pyx (Schaefer 2023b).  The size of the kink is quantified with the orbital periods immediately before and after the eruption are $P_{\rm pre}$ and $P_{\rm post}$.  The period change across the eruption is $\Delta P$=$P_{\rm post}$-$P_{\rm pre}$.  Steady period changes during quiescence can be notated as the time derivative of the period, $\dot{P}$.  The period change is a dimensionless quantity, or it could be stated as having units of seconds-per-second or days-per-day.  The $\dot{P}$ can change greatly across eruptions (see especially Schaefer \& Myers 2025), for unknown reasons, so I separate out $\dot{P}_{\rm pre}$ and $\dot{P}_{\rm post}$.  Here, the epoch, $E_0$, is allowed to be a free fit parameter, with this being close to the epoch from the ephemeris.

Now, I can run the chi-square fit between the 20 $T_{\rm min}$ measures and the model.  The best fit has $P_{\rm pre}$=0.1883896$\pm$0.0000018 days, $P_{\rm post}$=0.1884043$\pm$0.0000018 days, and $E_0$ equal to HJD 2459346.0601$\pm$0.0052.  These values are correlated, with a slightly late (early) $E_0$  making for a slightly larger (smaller) $P_{\rm pre}$ and for a slightly shorter (larger) $P_{\rm post}$, and hence a smaller (larger) $\Delta P$.  With these correlations, the best fit value of $\Delta P$ is $+$0.0000124 days, with a one-sigma range from (9.1--16.4)$\times$10$^{-6}$ days.  This makes the fractional change in the period ($\Delta P$/$P$) to be 66 ppm, with a one sigma range of 48--87 ppm.

The best-fitting $\dot{P}$ values are $\dot{P}_{\rm pre}$=($+$1.9$\pm$1.1)$\times$10$^{-9}$ and $\dot{P}_{\rm post}$=($-$1.2$\pm$1.7)$\times$10$^{-9}$.  The steady period change before the eruption is positive, while the steady period change after the eruption is consistent with zero.  

All 14 novae with well-measured $\dot{P}$ have values greatly different from zero, except for DQ Her (Schaefer 2023b).  So we strongly expect that V1405 Cas will have non-zero $\dot{P}$.  Nevertheless, if we assume that the $\dot{P}$ values are zero, then the best fit has $P_{\rm pre}$=0.1883955$\pm$0.0000004 days and $P_{\rm post}$=0.1884043$\pm$0.0000002 days, for $\Delta P$/$P$ to be 78 ppm. 

One important issue is whether the observed $\dot{P}$ values agree with the MBM.  This model requires that all the $\dot{P}$ values be small (for $P$=0.1884 days) and {\it negative}.  The measured values for V1405 Cas have relatively large uncertainty, because the time intervals in the $O-C$ curve are relatively small, both before and after the eruption.  The $\dot{P}_{\rm post}$ value is negative, but the error bar is large, so it is also consistent with zero or positive values.  The period changes before and after eruption are below the 3-sigma detection limits and consistent
with zero.  The $\dot{P}_{\rm pre}$ value is {\it positive}, but only at the 1.7-sigma confidence level.  So V1405 Cas provides a weak counterexample to a strong requirement of the MBM.  

Another period change of interest is the difference in $P$ from the earliest to the latest years over which we have data.  I do not have the measured $T_{\rm min}$ to directly calculate $P$ for 2013.0 or 2025.0, so I can only resort to the best fitting model.  With the model, the period for the year 2013.0 is 0.1883880 days, and the period for 2025.0 is 0.1884022 days.  That is, over all the years on observation, with all the effects included, the period of V1405 Cas {\it increased} by 0.0000142 days.  The fractional {\it increase} was $+$75 ppm.

\section{$M_{\rm ejecta}$}

\subsection{Old Methods for Measuring or Modeling $M_{\rm ejecta}$}

The purpose of this paper is to measure $M_{\rm ejecta}$ for V1405 Cyg by using a method from the  observed $\Delta P$.  But first, I should justify why this effort is critical.  After all, we could simply use any of the traditional methods for measuring $M_{\rm ejecta}$, or we could simply look up some theoretical calculation for $M_{\rm ejecta}$.  Unfortunately, the traditional measures and the theoretical estimates both have real total error bars typically between factors of 100$\times$ to 1000$\times$, making both useless for many purposes.  This might be surprising to some workers, so I need to provide a full analysis and justification.

For estimating $M_{\rm ejecta}$, both the traditional observational methods and the theory models have a real accuracy that is no better than a factor of 300$\times$ (see Appendix A of Schaefer 2011 and Sections 6.1 and 6.2 of Schaefer \& Myers 2025).  Here is a brief summary of the proofs that the errors are huge both for observational estimates and for theoretical estimates:  

{\bf Observational} estimates of $M_{\rm ejecta}$ are all with various traditional methods involving the equivalent of using hydrogen emission fluxes to derive the ejecta mass.  Unfortunately, these traditional methods all have five separate sources of systematic errors each at the 1-to-3 orders-of-magnitude level.  Problems include not knowing the gas temperature, the shell volume, and the filling factor, each creating orders-of-magnitude errors.  Together, all estimates of measured shell mass have 2--3 orders of magnitude uncertainty.  This result is confirmed by the 2--3 orders-of-magnitude scatter amongst independent estimates for individual novae\footnote{An example of critical importance here is for HR Del, where published measures from the traditional methods give ejected masses of 0.9 (Anderson \& Gallagher 1977), 2.5 (Malakpur 1973), 15 (Robbins \& Sanyal 1978), and 1.0--1.5 (Tylenda 1979), all in units of 10$^{-4}$ $M_{\odot}$.  These estimates span a range of 17$\times$.  For the important case of U Sco, published values of $M_{\rm ejecta}$ are 7.2 (Barlow et al. 1981), 1--10 (Williams et al. 1981), $\sim$10 (Anupama \& Dewangan 2000), 72--230 (Banerjee et al. 2010), and 210 (Pagnotta et al. 2015), all in units of 10$^{-8}$ $M_{\odot}$.  These estimates span a range of 230$\times$.  For the SD progenitor candidate T Pyx, published $M_{\rm ejecta}$ values based on the traditional methods are 10--300 (Nelson et al. 2014), $\sim$400 (Caleo \& Shore 2015), 7.03 (Pavana et al. 2019), 10--100 (Selvelli et al. 2008), 30 (Schaefer et al. 2010), $\gtrsim$10 (Chomiuk et al 2014), and $<$(3$\pm$1) (Izzo et al. 2024), all in units of 10$^{-6}$ $M_{\odot}$.  These estimates span a range of $>$130$\times$.  This record of huge scatter is the proof that observational estimates of $M_{\rm ejecta}$ have real uncertainties of 2--3 orders-of-magnitude.  }.  

{\bf Theoretical} models for estimating $M_{\rm ejecta}$ have three huge problems.  One primary problem is that the calculated ejecta masses are sharply sensitive to innocuous and unknowable details in the model calculation\footnote{Kato, Hachisu, \& Saio (2017) chronicle the bickering over the computational details.  Critical problems include the number of layers in the mass grids, the number of flash cycles included in the calculation, whether helium flashes are included, and the mass loss algorithm.  With this, there is no wonder as to why modelers report a huge range of $M_{\rm ejecta}$ for identical cases.  Starrfield et al. (2020) give a study where one parameter in their code is changed over a small range, and the calculated $M_{\rm ejecta}$ is reported.  In their Appendix Table A1, the $M_{\rm ejecta}$ is reported for the cases of using 95, 150, 200, and 300 mass zones, with ejections of 0.24, 3.6, 1.9, and 5.7 in units of 10$^{-8}$ $M_{\odot}$.  There is no indication of convergence, and the systematic errors from the simple choice of the number of mass zones to be used in the code causes a variation of 24$\times$.  It is horrifying that the published $M_{\rm ejecta}$ varies so hugely on just one of many arbitrary choices in programming.}.  The second primary problem is that the calculated ejecta masses are sharply sensitive to unknown properties of the WD\footnote{Starrfield et al. (2025) give a study where one parameter is changed at a time, and the calculated $M_{\rm ejecta}$ is reported.  For the T CrB case of a 1.35 $M_{\odot}$ ONe WD, the ejecta mass changes from 5.3 to 10.7 to 1.2 (in units of 10$^{-8}$ $M_{\odot}$) as the WD radius changes from 1522 to 1827 to 2166 km.  That is a factor of 9$\times$ just for not knowing the WD size.  In changing from an assumed oxygen fraction in the burning gas of 25\% to 50\%, for the middle WD radius, the ejecta mass changes from 1.5 to 10.2 times 10$^{-8}$ $M_{\odot}$.  This is a factor of 7$\times$ systematic error for the unknowable oxygen fraction.  And if the WD is taken to be of CO composition, then the ejecta mass goes up to 35.1 times 10$^{-8}$ $M_{\odot}$.  So usually unknowable WD properties make for systematic errors of up to 29$\times$.  Starrfield et al. (2024) report single parameter variations for a model with WD mass of 0.60 $M_{\odot}$ (as appropriate for V1405 Cas) where the small changes in the composition of the surface layer make for 2$\times$ changes in $M_{\rm ejecta}$, and varying the WD mass within the stated error bars changes $M_{\rm ejecta}$ by a factor of 4$\times$.  Further, simply by changing the unknowable mixing from dredge-up during the thermonuclear runaway, the $M_{\rm ejecta}$ changes from 49 to 16,000 in units of 10$^{-8}$ $M_{\odot}$, for a systematic error of 300$\times$.  In all, ordinary free choices of unknown WD properties makes for real uncertainties of $>$300$\times$.}.  The first two problems are confirmed when we see huge scatter of published models for the same individual novae\footnote{For U Sco, published theory estimates of $M_{\rm ejecta}$ are 21 (Kato 1990), $\sim$180 (Hachisu et al. 2000), 43 (Starrfield et al. 1988), 440 (Yaron et al. 2005), plus 110 and 210 (Figueira et al. 2025), all in units of 10$^{-8}$ $M_{\odot}$.  This covers a range that is a factor of 21$\times$ wide.  This theory range has only modest overlap with the observational range (see footnote 9), and the entire range of estimates for U Sco is from 1--440 $\times$10$^{-8}$ $M_{\odot}$, for a total range over a factor of 440$\times$.  Starrfield et al. (2020) reports on four independent calculations of $M_{\rm ejecta}$ for identical cases by different groups.  For a typical comparison for a 1.15 $M_{\odot}$ WD, they collect ejecta masses of 1500, 490, 0.98, and 1300 (in units of 10$^{-8}$ $M_{\odot}$).  This is a range of 1530$\times$.  This is proof that modern theory estimates of $M_{\rm ejecta}$ have real error bars of three orders-of-magnitude in size.}.  The third primary problem is that all the old models are just the pedestrian 1-dimensional models generating optically thick winds, with these all completely omitting what we now know to be the dominant mass ejection mechanism, so all prior model estimates of $M_{\rm ejecta}$ are now repudiated.  

These observational and theoretical problems were well known, most even back in 1983, with this explicitly providing a required part of the motivation for measuring $\Delta P$ for BT Mon as a dynamical measure of $M_{\rm ejecta}$ (Schaefer \& Patterson 1983).  For purposes of this paper, we must acknowledge and use error bars of 2--3 orders-of-magnitude for application to HR Del and RR Pic.

\subsection{$\Delta P_{\rm ejecta}$}

The key relation here is that the binary must change its period when its nova ejects gas from the WD.  With the observed $\Delta P$, we can work backwards to calculate $M_{\rm ejecta}$.  To start, the observed $\Delta P$ is the sum of two effects, the period change forced by the mass loss ($\Delta P_{\rm ejecta}$) and the period change forced by angular momentum loss (`aml') during the nova eruption ($\Delta P_{\rm aml}$).  So
\begin{equation}
\Delta P = \Delta P_{\rm ejecta} + \Delta P_{\rm aml}.
\end{equation}
For all binaries for all cases, 
\begin{eqnarray}
\Delta P_{\rm ejecta}>0, \nonumber \\
\Delta P_{\rm aml}<0.
\end{eqnarray}
These constraints are because the binary cannot gain mass from the outside and cannot gain angular momentum from the outside.  So whether the binary has a positive or negative $\Delta P$ depends on a balance between the mass loss and the angular momentum loss of the binary.  If $\Delta P$ is positive (as for V1405 Cas), then the mass loss dominates over the angular momentum loss, and this requires some relatively large $M_{\rm ejecta}$.

The relation between $M_{\rm ejecta}$ and $\Delta P$ can be derived easily from Kepler's Law and the conservation of angular momentum.  This derivation was widely known long before 1983 (e.g., Ahnert 1960, Huang 1963), when it became a primary motivation for measuring the period change of BT Mon.
\begin{equation}
\Delta P_{\rm ejecta} = 2  \frac{M_{\rm ejecta}}{M_{\rm comp}+M_{\rm WD}} P.
\end{equation}
With $P$, $M_{\rm comp}$, and $M_{\rm WD}$ known with usable accuracy, we have a simple and direct means to calculate $M_{\rm ejecta}$ with useable accuracy.  

The use of Equation 5 has some great advantages.  As a simple physics result derived only with Kepler's Law and the conservation of angular momentum, there is no chance of any loopholes, exceptions, or alternatives.  The input physics is based on conservation laws applied to the states before and after the eruption, so all the complications during the nova do not enter the problem.  The application of Equation 5 is a simple timing experiment, for which high-accuracy can be produced by the timing of many eclipses across 23,286 orbital cycles of V1405 Cas.   The input and analysis only involves eclipse minimum timings, which is a tremendous relief because the total error bars are simple, with none of the huge systematic problems that plague the traditional methods (distance, extinction, filling factors, gas temperatures, ionization states, and the wide range of gas densities).  Hence, the $\Delta P_{\rm ejecta}$ method is accurate and reliable.   

Equation 5 is the basis for my career-long program of measuring $M_{\rm ejecta}$ for nova eruptions.  This method is needed because all the traditional methods (involving measures of hydrogen line fluxes) have real uncertainties of two-to-three orders of magnitude (see Section 4.1).  Unlike the traditional methods, the $\Delta P$ method for measuring $M_{\rm ejecta}$ has no dependency on extinction, distance, gas temperature, filling factors, any of the physics of the explosion, or even on the nature of the ejection mechanism.  Rather, the $\Delta P$ method is an experiment based on simple dynamics.  Rather, the $\Delta P$ method is just a simple timing experiment.  As such, with an adequate set of eclipse times, the mass of the ejecta can be reliably measured with good accuracy.  

The only issue is that the possibility of angular momentum loss during the eruption will make the derived mass as a limit (see Equations 3-5).  If the angular momentum loss is negligibly small, then Equation 5 will give the mass of the ejecta.  If there is any substantial angular momentum loss, then $M_{\rm ejecta}$ will only be {\it larger} than calculated from Equation 5.  

Is there some sort of an upper limit on the $M_{\rm ejecta}$?  Hachisu \& Kato (2026) claim that there is an upper limit, based on the assumption that all of the energy for ejecting mass comes solely from the fast thermonuclear runaway (TNR) that starts the explosion.  The trouble is that this is an old now-disproven model assumption, whereas many recent workers (Chomiuk, Metzger, \& Shen 2020, Sparks \& Sion 2021, Shen \& Quataert 2022, Schaefer 2026a, Mukai 2026) have realized that there are actually large energy sources not considered in the old-style theory.  This started with much accumulating observations that showed the nova mass ejections are long and drawn out, lasting months to years.  Such is impossible within the old-style 1-dimensional theory where the only energy source is the TNR that lasts bare minutes.  The late ejections dominate, and these must have some large energy source(s) not considered by the old-theory.  Two energy sources have been considered, with both being inevitable.  One extra energy source is the companion's orbital energy, where the companion stirs up and ejects massive gases from the hot supersoft envelope that surrounds the exploding WD for the next many months.  Another extra energy source is the gravitational potential energy of the supersoft envelope, which must be released over its lifetime, where such fallback will drive a powerful disk wind at the Eddington rate.  The situation is much too complex for current models to predict $M_{\rm ejecta}$, but the usual approximations shows that these two `new' energy sources are large, much larger than the TNR alone.  With this, there is no upper limit on $M_{\rm ejecta}$.

From above, $M_{\rm comp}$=0.43$\pm$0.04 $M_{\odot}$ and $M_{\rm WD}$=0.60$\pm$0.10 $M_{\odot}$.   To frame the reasonable range, an ejecta mass of 10$^{-5}$ $M_{\odot}$ makes for  $\Delta P_{\rm ejecta}$/$P$ to equal 20 ppm, while an ejecta mass of 10$^{-4}$ $M_{\odot}$ makes for  $\Delta P_{\rm ejecta}$/$P$ to equal 200 ppm.

\subsection{Old $\Delta P_{\rm aml}$ Mechanism}

The angular momentum loss mechanism during a nova eruption is the so-called frictional angular momentum loss.  This dynamical AML arises from the companion star dragging through the gas thrown out by the nova eruption.  The situation is that of a common-envelope, where the dynamical drag (like Bondi-Hoyle accretion) of the gas on the companion star will create a force that slows down the companion's orbital velocity.  This will force the companion into a smaller orbit with a shorter period.  With a smaller orbit, the companion has smaller orbital angular momentum.  This lost angular momentum can only go into the gas, which is then ejected from the system.  How much angular momentum is lost depends on the details of the gas thrown off by the nova eruption.

Historically, the first estimate of $\Delta P_{\rm aml}$ was in Livio, Govarie, \& Ritter (1991).  They made the reasonable model that the dynamical friction was operating with all of the ejecta gas traveling outward at the velocity of the ejecta ($V_{\rm ejecta}$).  This would be the case if all the ejecta were blasted off the WD by the initial thermonuclear runaway.  If all the ejected gas passes by the companion at high velocity, then
\begin{equation}
\Delta P_{\rm aml} = -\frac{3}{4} P \frac{M_{\rm ejecta}}{M_{\rm comp}} \frac{V_{\rm orb}}{V_{\rm shell}} \left( \frac{R_{\rm comp}}{a} \right)^2.
\end{equation}
$V_{\rm orb}$ is the usual Keplerian velocity of the companion star in its orbit.  $V_{\rm shell}$ is the expansion velocity measured by the FWHM of the spectrum or by resolving the expanding shell.  The companion star's radius is $R_{\rm comp}$ and $a$ is the semimajor axis of the orbit.  From equations 5 and 6, $\Delta P_{\rm aml}$ will always be negative, and $|\Delta P_{\rm aml}|$ will always be greatly smaller than $\Delta P_{\rm ejecta}$.  That is, with the original formulation of AML, the mass loss from the ejecta always dominates, so the observed $\Delta P$ should always be positive.  This formulation of AML has been the mainstay of nova evolution calculations up until a few years ago.

For the stellar masses from above, the semimajor axis is 1.38 $R_{\odot}$, $R_{\rm comp}$=0.49 $R_{\odot}$, and $V_{\rm orb}$=220 km/s.  $V_{\rm shell}$ is best estimated from the median FWHM for hydrogen lines of 1250 km/s.  Just to set the scale, an ejecta mass of 10$^{-5}$ $M_{\odot}$ makes for  $\Delta P_{\rm aml}$/$P$ to equal $-$0.7 ppm, while an ejecta mass of 10$^{-4}$ $M_{\odot}$ makes for  $\Delta P_{\rm aml}$/$P$ to equal $-$7 ppm.  Given the factors in Equations 5 and 6, the ratio $\Delta P_{\rm ejecta}$/$\Delta P_{\rm aml}$ equals $-$29.  This is to say that this part of the AML effect is small.

\subsection{The New $\Delta P_{\rm aml}$ Mechanism That Dominates}

In recent years, various workers (Chomiuk, Metzger, \& Shen 2020, Sparks \& Sion 2021, Shen \& Quataert 2022) have discovered that the mechanism in Equation 6 is dominated by a different AML mechanism with greatly larger effect\footnote{The realization is that the long standard picture (a near-Eddington wind launched from near the WD) cannot work for most novae because the wind launch zone is out past the binary orbit, so this manner of ejection is largely cutoff by interference from the companion.  Shen \& Quataert (2022) explain: ``Crucially, we find that, even for successful winds, the velocity at the location of the WD's Roche radius is slower than the typical orbital velocities of WDs in CVs for most of the mass-loss phase, because the acceleration region is outside of the WD's Roche lobe. As a result, standard optically thick winds cannot form, and binary interaction will instead initiate the majority of mass loss, focusing it toward the equatorial plane.''}.  In particular, the companion suffers large dynamical drag as it passes through the outer parts of the envelope puffed up around the white dwarf as part of the nova eruption.  This hot envelope will appear as the supersoft X-ray source later in the eruption.  For most novae, the envelope extends densely out to the companion's orbit, where the envelope/companion interaction serves to eject most of the mass.  There is nuclear burning on-going near the WD surface, and this powers the expansion of the envelope, and when this burning runs out, the supersoft source turns off, the envelope collapses, and the mass ejection stops.

Observational evidence for this paradigm is presented in the review by Chomiuk, Metzger, \& Shen (2020).  Spectroscopic evidence points to frequent equatorial rings/disks and bipolar shapes for the outflow, demonstrating that the companion and its orbit are shaping, if not driving, the mass ejection.  Such cannot be explained by any impulsive fast ejection, nor from any prolonged optically-thick winds.  Pictures of nova shells (Santamar\'{i}a et al. 2025) show that the slower novae have the most elongated shapes (Slavin et al. 1995, O'Brien \& Bode 2008), so slow novae like V1405 Cas will have the mass ejection dominated by the binary interaction with the hot envelope.  For the well-observed RN T Pyx, Chomiuk et al. (2014) find that the shell was ejected in two phases, ``one shell is expelled on Day 0 with a velocity of $\sim$1900 km/s and a second is expelled later on with a velocity of 3000 km/s The bright hard X-ray component appears in the Swift data on Day 117, and rises steeply until it reaches maximum on Day 142 and plateaus until Day $\sim$206, after which the flux gradually fades.''  That is, the initial blast is from the short-duration of the thermonuclear runaway, which carries only a small fraction of the ejected gas, while the subsequent shell is by the mechanism of the binary/envelope ejection, which lasts for longer than 200 days.  My favorite evidence is my measure of the timing of the kink in the $O-C$ curve of the 2011 eruption of T Pyx, where the {\it median} epoch of mass ejection was 108.4$\pm$12.6 days after the start of the eruption, by which time the nova had faded by 5 mags, proving that the ejection was a long affair lasting several hundred days (Schaefer 2023b).

Unfortunately, this new mechanism cannot yet be accurately modeled with theory.  There are two major problems.  First, the structure of the hot envelope is poorly known, even for the case with the WD being isolated.  The only modeling that I have seen is in Shen \& Quataert (2022), this is only 1-dimensional with a variety of effects missing. Second, and most importantly, the interaction of the companion with the hot envelope is a complex 3-dimensional calculation, involving the companion sculpting the envelope in currently un-knowable ways, with such sculpting making for large changes in the modeled $M_{\rm ejecta}$ and $\Delta P_{\rm aml}$.  It appears that our community is far away from being able to model even an order-of-magnitude estimate of $\Delta P_{\rm aml}$. 

Even without a detailed 3-dimensional physics model, we already know some of the primary scaling factors.  In particular, the AML effect will be proportional to the gas density in the hot envelope at the position of the companion's orbit, and proportional to the duration of the hot envelope.  

The density in the WD hot envelope falls off roughly as $r^{-3}$, where $r$ is the radial distance from the WD (Shen \& Quataert 2022).  With Kepler's Law, this makes the gas density along the companion star's orbit scale as $M_{\rm WD}^{-1}P^{-2}$.  Further, Shen \& Quataert (2022, see their figure 11) calculate that the mass in the envelope (and its density) scales as $10^{-2.5 M_{\rm WD}}$.  

The AML effect is also proportional to the duration of the hot envelope, with this duration being roughly the duration of the supersoft phase of the nova eruption.  The supersoft phase is only measured for novae in the last three or four decades, so for earlier novae (like for the twins HR Del and RR Pic, see below), we need some proxy for the turnoff time.  The proxy could be either the $t_2$ or the $t_3$, or the FWHM of the Balmer lines (see Schwarz et al. 2011).  For a proxy usable with old and new novae, the best is apparently the correlation fit from Greiner, Orio, \& Schartel (2003), which has the supersoft turnoff time proportional to $FWHM^{-2.1}$, as confirmed by Schwarz et al. (2011).  Schaefer (2025d) find a relation where the FWHM is proportional to $10^{M_{\rm WD}/2}$, with the WD mass in solar units.  This then gives the AML effect as being proportional to $10^{-M_{\rm WD}}$, to appropriate precision.  

So, with two of the dominating effects accounted for, we have 
\begin{equation}
\frac{\Delta P_{\rm aml}}{P} = -c P^{-3} M_{\rm WD}^{-1} 10^{-3.5 M_{\rm WD}} .
\end{equation}
Here, $c$ is a positive scaling constant, $P$ is in units of days, and $M_{\rm WD}$ is in units of $M_{\odot}$.  All of the input relations display considerable scatter, indicating that conditions are complex with other unresolved factors, as expected.  Undoubtedly, further trends and mechanisms can be included to fine-tune Equation 7.  So Equation 7 does not have any high accuracy.  Nevertheless, Equation 7 is adequate for seeing the dominant effects, and how they vary system-to-system.

We can see how this works for various novae.  If the period is long and the WD mass is high, then the $\Delta P_{\rm aml}$ will be negligibly small.  For the case of U Sco, with $P$=1.23 days and $M_{\rm WD}$=1.37 $M_{\odot}$, $\Delta P_{\rm aml}$ is $-$0.000008 (with $c$ set to unity).  If the period is short and the WD mass is low, then $\Delta P_{\rm aml}$ will be large and negative.  For the case of V1405 Cas, with $P$=0.1884 days and $M_{\rm WD}$=0.60 $M_{\odot}$, $\Delta P_{\rm aml}$ is $-$0.37 (again with $c$=1).  That is, we expect that V1405 Cas to have a 48,000$\times$ larger period change from AML than for U Sco\footnote{For U Sco, the supersoft source turned off after 34$\pm$4 days.  For V1405 Cas, the supersoft source lasted 1800 days.  If we take the observed durations from the turnoff times, and apply the scaling for the density, then we get an expected ratio of 430,000$\times$.  This order of magnitude difference is notice that Equation 7 has only order-of-magnitude accuracy.}.  The important point for this paper is that $\Delta P_{\rm aml}$ is large and negative for V1405 Cas.  Indeed, with V1405 Cas having one of the smallest-mass WDs and a short $P$, this nova should have one of the largest AML period decreases out of all novae.

\subsection{Constraints on $\Delta P_{\rm aml}$}

\begin{table*}
	\centering
	\caption{Properties of the triplets HR Del, RR Pic, and V1405 Cas}
	\begin{tabular}{lrrr}
		\hline
		   &  HR Del  &   RR Pic   &  V1405 Cas \\
		\hline
$P$ (days)	&	0.2142	&	0.1450	&	0.1884	\\
Light curve class (SPOCJDF)	&	J	&	J	&	J	\\
$t_3$ (days)	&	231	&	122	&	175	\\
Spectral class	&	Fe II	&	Fe II	&	Hybrid$^a$	\\
WD composition	&	ONe	&	ONe	&	ONe	\\
$M_{\rm WD}$ ($M_{\odot}$)	&	0.67	&	0.66	&	0.60	\\
$M_{\rm comp}$ ($M_{\odot}$)	&	0.51	&	0.23	&	0.43	\\
$\dot{M}$ ($M_{\odot}$ yr$^{-1}$)	&	$2 \times 10^{-8}$	&	$2 \times 10^{-8}$	&	$1 \times 10^{-8}$	\\
$\Delta P$/$P$ as observed (ppm)	&	$-$472.1	&	$-$2003.7	&	$+$66	\\
$M_{\rm ejecta}$ adopted (10$^{-4}$ $M_{\odot}$)	&	0.1 to 30	&	0.1 to 30	&	...	\\
$\Delta P_{\rm ejecta}$/$P$ limit (ppm)	&	$+$17 to $+$5100	&	$+$22 to $+$6700	&	...	\\
$\Delta P_{\rm aml}$/$P$ limit (ppm)	&	$<$ $-$472.1	&	$<$ $-$2003.7	&	...	\\
$\Delta P_{\rm aml}$/$P$ best (ppm)	&	$-$640	&	$-$2200	&	$\Rightarrow$~$-$1400	\\
$\Delta P_{\rm aml}$/$P$ range (ppm)	&	$-$5600 to $-$490	&	$-$8700 to $-$2000	&	$\Rightarrow$~$-$8700 to $-$490	\\
$\Delta P_{\rm ejecta}$/$P$ best (ppm)	&	...	&	...	&	$+$1400	\\
$\Delta P_{\rm ejecta}$/$P$ range (ppm)	&	...	&	...	&	$+$490 to $+$8700	\\
$M_{\rm ejecta}$ best ($M_{\odot}$)	&	...	&	...	&	7.5$\times$$10^{-4}$	\\
$M_{\rm ejecta}$ range ($M_{\odot}$)	&	...	&	...	&	(2.9--40)$\times$$10^{-4}$	\\
		\hline
	\end{tabular}	
	\\$^a$V1405 Cas is officially a hybrid spectral type, \\yet most of the days in the eruption were spent as spectral class Fe II.	
\end{table*}

$\Delta P_{\rm aml}$ cannot yet be calculated from first principles, yet we nevertheless have a number of constraints.  The AML effect must be negative, with the orbital period decreasing, because there is no outside source to add angular momentum to the binary.  Another constraint is that we know the AML effect is amongst the largest (negative) values for all novae.  Further constraints can be gotten from two novae that are close twins of V1405 Cas.

For nova eruption physics and binary properties, the most important are the orbital period $P$, the light curve class, the nova light curve timescale $t_3$, the spectral class, the WD composition, the mass of the WD $M_{\rm WD}$, the mass of the companion star $M_{\rm comp}$, and the accretion rate $\dot{M}$.  Largely, all the eruption physics is determined and expressed by just these parameters\footnote{We can conceive of other parameters that might effect the physics of the nova eruption, for which I expect that the WD magnetic field may have some observable implications.  Indeed, U Sco has the large eruption-to-eruption changes in the five observed $\Delta P$ measures, despite all its nova eruptions being indistinguishable as seen with extensive photometry and spectroscopy.  }.  Fortunately, V1405 Cas has two near-perfect twins, HR Del and RR Pic.  Maybe we should call the three twins as a triplet.  HR Del and RR Pic are the 11th and 1st brightest novae in the last century, and are some of the best observed.  Fortunately, both are ones that I have measured $\Delta P$.

Table 2 lists the eight principle properties of the triplet.  All of these properties are essentially identical, as compared to the range of nova properties.  For the orbital periods, the triplets are not identical, but they are all somewhat above the Period Gap, so operationally they have the same physics and evolution.  For the $t_3$ measures, they are not identical, but all three are amongst the slowest novae, so they have the same physics and case for the eruption.  For all eight primary properties, V1405 Cas, HR Del, and RR Pic are startlingly similar.

We can test the similarity of this triplet for $\Delta P_{\rm aml}$/$P$ by using Equation 7.  Ratios of the model $\Delta P_{\rm aml}$/$P$ can be calculated without needing to know a value for $c$.  The ratio for V1405 Cas to HR Del is 2.9, while the ratio from V1405 Cas to RR Pic is 0.8.  With the uncertainty on $M_{\rm WD}$ for V1405 Cas alone, the error bars shows the ratios to be consistent with unity.  That is to say, to within the error bars, the $\Delta P_{\rm aml}$/$P$ values for three stars in the triplet are identical.  The range of ratios for the triplet are greatly smaller than the range for novae.  The reason is that Equation 7 shows that the dependencies on $P$ and $M_{\rm WD}$ are both very strong.  For novae with periods ranging from 0.1 days to 1.0 days, the ratio changes by a factor of 1000$\times$.  For novae with WD masses ranging from 0.6 to 1.3 $M_{\odot}$, the ratio changes by a factor of 610$\times$.  With these huge variations, the differences within the triplet are tiny.  This further justifies the three novae as being identical triplets, to within the known uncertainties.  And this is further justification that the measured $\Delta P_{\rm aml}$/$P$ values for HR Del and RR Pic are closely applicable to V1405 Cas.

Both HR Del and RR Pic have negative $\Delta P$.  These two twins have by-far the most extreme-negative $\Delta P$ values ever measured, and this is likely related to the twins also having extremely low WD masses.  With this, we can place useful limits on $\Delta P_{\rm aml}$.  From Equation 3 and 4, we know $\Delta P_{\rm aml}$$<$$\Delta P$.  For those novae with negative $\Delta P$, this limit is useful.  Table 2 lists these limits on $\Delta P_{\rm aml}$/$P$ for HR Del and RR Pic.  The two limits ($<$ $-$472.1 and $<$ $-$2003.7 ppm) for the HR Del and RR Pic twins should also be applicable to V1405 Cas.  

These limits are calculated for assumed zero-$M_{\rm ejecta}$, and that is why they are limits.  Fortunately, we do have some measures of $M_{\rm ejecta}$ for HR Del and RR Pic, and these can be used to calculate $\Delta P_{\rm aml}$.  Importantly, we must realize that all the published measures of $M_{\rm ejecta}$ are highly uncertain (see Section 4.1), so we must take this to heart and apply appropriately large error bars of the averaged measured $M_{\rm ejecta}$ values.

I have collected published estimates of $M_{\rm ejecta}$ for HR Del and RR Pic.  For HR Del, we have four reports of shell masses with the traditional methods.  These report 0.9 (Anderson \& Gallagher 1977), 2.5 (Malakpur 1973), 15 (Robbins \& Sanyal 1978), and 1.0--1.5 (Tylenda 1979), all in units of 10$^{-4}$ $M_{\odot}$.  For RR Pic, we only have the mass from Celed{\'o}n et al. (2024) at 0.5 in units of 10$^{-4}$ $M_{\odot}$.  Yaron et al. (2005) present generic models that are applicable to HR Del and RR Pic.  These are for WD mass of 0.65 $M_{\odot}$ and accretion rates of 10$^{-8}$ $M_{\odot}$ yr$^{-1}$.  They calculate that $M_{\rm ejecta}$ is 1.0$\times$10$^{-4}$ $M_{\odot}$.  So for both of the twin-novae, the best overall estimate is something like 1.0$\times$10$^{-4}$ $M_{\odot}$.  However, taking to heart the huge real uncertainty, we can take the full range of estimates and expand this range by half an order-of-magnitude.  With this, the ejecta mass is (0.1--30)$\times$10$^{-4}$ $M_{\odot}$, for both HR Del and RR Pic.

With an estimate for the ejecta mass of HR Del and RR Pic (even with large error bars), we can get an estimate of $\Delta P_{\rm ejecta}$, and then an estimate of $\Delta P_{\rm aml}$.  This can be calculated as 
\begin{equation}
\frac{\Delta P_{\rm aml}}{P} =  \frac{\Delta P}{P}  -2  \frac{M_{\rm ejecta}}{M_{\rm comp}+M_{\rm WD}}.
\end{equation}
The best value for $\Delta P_{\rm aml}$/$P$ comes for the estimate of the ejecta mass of 1.0$\times$10$^{-4}$ $M_{\odot}$.  For HR Del, this best estimate is $-$640 ppm, with a full range from $-$5600 to $-$490 ppm.  For RR Pic, this best estimate is $-$2200 ppm, with a full range from $-$8700 to $-$2000 ppm.

We now have two values, with large ranges, that are measures of $\Delta P_{\rm aml}$/$P$ for novae that are near-perfect twins of V1405 Cas.  The average of the two best values ($-$640 and $-$2200 ppm), the best combined value for application to V1405 Cas is $-$1400 ppm.  The extreme range is $-$8700 to $-$490 ppm.  These ranges are illustrated graphically on a number line in Figure 4.  This is like us using the HR Del and RR Pic binaries as enormous analog computers for the extremely complex calculation of $\Delta P_{\rm aml}/P$, with the input appropriate for properties of V1405 Cas.

\begin{figure*}
	\includegraphics[width=2.01\columnwidth]{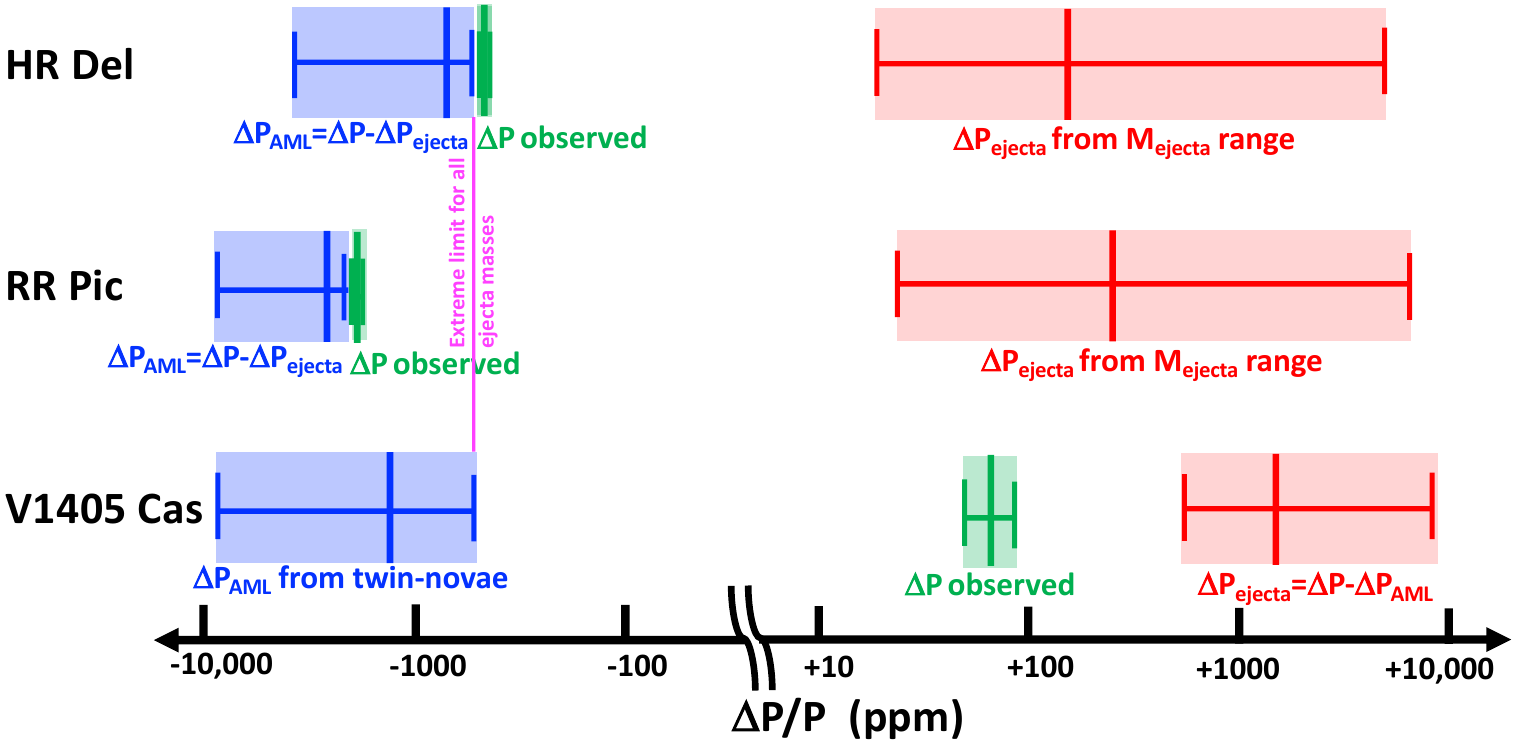}
    \caption{$\Delta P_{\rm aml}$/$P$ for V1405 Cas.  This number line for $\Delta P$/$P$ is on a logarithmic scale (in units ppm), with a necessary break around zero.  The three rows are graphic displays for HR Del (top row), RR Pic (middle row), and V1405 Cas (bottom row), showing the best values and ranges for $\Delta P$/$P$ (the green ranges), $\Delta P_{\rm aml}$/$P$ (the blue ranges), and $\Delta P_{\rm ejecta}$/$P$ (the red ranges).    The red ranges must necessarily be positive, while the blue ranges must necessarily be negative.  With $\Delta P$=$\Delta P_{\rm aml}$+$\Delta P_{\rm ejecta}$, the green ranges are a balance between the positive effects of the mass ejection as compared to the negative effects of the AML.  After the measures of $\Delta P$ for each nova, the analysis starts with evaluating $\Delta P_{\rm ejecta}$ as based on an extremely wide range (over a factor of 300$\times$ in width) of possible ejecta masses, expanded greatly from the various independent estimates for HR Del and RR Pic.  The $\Delta P_{\rm aml}$ is then calculated for both HR Del and RR Pic.  This knowledge from the two twin novae is applied to makes limits for V1405 Cas.  With $\Delta P$ and $\Delta P_{\rm aml}$ for V1405 Cas, we can calculate $\Delta P_{\rm ejecta}$ and $M_{\rm ejecta}$ for V1405 Cas.  This many step procedure is tedious to follow.  Fortunately, for the purposes of this paper, we can follow a simple and easy path to get a confident limit that answers the big-picture question.  That is, we know from the two twins (HR Del and RR Pic) that $\Delta P_{\rm aml}$ for V1405 Cas is more-negative than $-$490 ppm (to the left of the magenta line in the figure), which then forces V1405 Cas to have $M_{\rm ejecta}$$>$2.8$\times$$10^{-4}$ $M_{\odot}$.  With the extreme limit from this one-sentence method (and the accreted mass from the next Section), we have our answer that $M_{\rm ejecta}$$>$$M_{\rm accreted}$ and hence the WD in V1405 Cas is {\it decreasing} in mass over time, and hence it cannot possibly become a SNIa.   }
\end{figure*}

\subsection{The $\Delta P$ Method for $M_{\rm ejecta}$}

Now we have the observed $\Delta P$ and we have a real measure for $\Delta P_{\rm aml}$, so we can simply calculate the ejecta mass.
\begin{equation}
M_{\rm ejecta} = 0.5 (M_{\rm WD} + M_{\rm comp}) \left( \frac{\Delta P}{P} - \frac{\Delta P_{\rm aml}}{P} \right).
\end{equation}
With $\Delta P_{\rm aml}$/$P$ equaling $+$66 ppm and the best $\Delta P_{\rm aml}$/$P$ equaling $-$1400 ppm, then $M_{\rm ejecta}$ equals 0.00075 $M_{\odot}$.  So there we have it, the best estimate for the measured ejecta mass.

The uncertainty in this measure is almost entirely from the uncertainty in $\Delta P_{\rm aml}$/$P$.  For the extreme range of $-$8700 to $-$490 ppm, the $M_{\rm ejecta}$ varies from 0.00029 to 0.0044 $M_{\odot}$.  So, the best estimate is that $M_{\rm ejecta}$=7.5$\times$10$^{-4}$ $M_{\odot}$, with the extreme range being (2.9--40)$\times$10$^{-4}$ $M_{\odot}$.  That is, in all possible cases, $M_{\rm ejecta}$ $>$ 2.9$\times$10$^{-4}$ $M_{\odot}$.

\subsection{The Simple and Sure Limit On $M_{\rm ejecta}$}

Sections 4.4 to 4.6 (including Table 2 and Figure 4) give a number-ridden formal evaluation of $M_{\rm ejecta}$, with all the detailed calculations effectively hiding the analysis for the readers that do not spend the time to look closely.  Fortunately, there is a simple and sure way to put a confident limit on $\Delta P_{\rm aml}$ and then $M_{\rm ejecta}$.  The key point is that the two twin novae give us a confident limit that $\Delta P_{\rm aml}$/$P$ is $<$$-$472 ppm.  That is, even if HR Del and RR Pic have near-zero ejecta mass, the $\Delta P_{\rm aml}$ must be more negative than the observed $\Delta P$.  With this extreme case, $\Delta P_{\rm ejecta}$/$P$ for V1405 Cas is always more negative than $+$66 ppm minus $-$472 ppm, which is to say that the extreme possible lowest value for V1405 Cas is $+$538 ppm.  With $\Delta P_{\rm ejecta}$$>$538 ppm, then Equation 5 yields $M_{\rm ejecta}$$>$2.8$\times$$10^{-4}$ $M_{\odot}$.  So we have a simple and confident extreme limit, and this limit is all we need to answer the big question of whether $M_{\rm ejecta}$ is larger or smaller than $M_{\rm accreted}$.

\section{$M_{\rm accreted}$}

For CV evolution, a critical question is the balance between the mass ejected by the white dwarf during each eruption ($M_{\rm ejecta}$) and the mass accreted by the white dwarf during the prior inter-eruption interval ($M_{\rm accreted}$).  $M_{\rm accreted}$ can be calculated as the average $\dot{M}$ times the recurrence timescale ($\tau_{\rm rec}$).  $M_{\rm accreted}$ is also equal to the mass of the accumulated gas required to trigger the runaway thermonuclear explosion, $M_{\rm trigger}$.  We have a reasonable measure of $\dot{M}$, but no {\it observational} basis for $\tau_{\rm rec}$ or $M_{\rm trigger}$.  Fortunately, we have fairly good {\it theoretical} knowledge for estimating $\tau_{\rm rec}$ and $M_{\rm trigger}$.

Yaron et al. (2005) has set up a large grid of models, with the needed calculations.  For a 0.65 $M_{\odot}$ WD that accretes at 10$^{-8}$ $M_{\odot}$ yr$^{-1}$, the recurrence timescale is 10,100 years and hence $M_{\rm accreted}$ is 1.0$\times$10$^{-4}$ $M_{\odot}$.  For interpolating to 0.60 $M_{\odot}$, the model $M_{\rm accreted}$ is 1.3$\times$10$^{-4}$ $M_{\odot}$.

Shen \& Bildsten (2008) report on $\tau_{\rm rec}$ and $M_{\rm trigger}$ for a large grid covering a wide range of $\dot{M}$ and $M_{\rm WD}$.  For a 0.6 $M_{\odot}$ WD accreting at a rate of 10$^{-8}$ $M_{\odot}$ yr$^{-1}$, $\tau_{\rm rec}$ is 20,000 years and $M_{\rm trigger}$ is 2.0$\times$10$^{-4}$ $M_{\odot}$.

These calculations are solid and well-known physics, so the differences between the two theory calculations are likely just from differing cases, for example, due to the WD and the metallicity.  So all I can do is to take the middle, and have the differences give the error bar.  So I am concluding that $M_{\rm accreted}$=$M_{\rm trigger}$=(1.6$\pm$0.4)$\times$10$^{-4}$ $M_{\odot}$.

\section{Supernova progenitor?}

The original idea back in 1983 was to test whether $M_{\rm ejecta}$$>$$M_{\rm accreted}$ to answer the question whether novae are the progenitors of Type Ia supernovae (SNIa).  Since then, a second test is whether the WD is of CO or ONe composition.  Now, we have enough information to apply both tests to V1405 Cas.

The main task of this paper is to measure the orbital period changes and then calculate $M_{\rm ejecta}$.  In Section 4.3, I found that the best estimate ejecta mass is 7.5$\times$10$^{-4}$ $M_{\odot}$, while there is a confident extreme limit that the ejecta mass is $>$2.9$\times$10$^{-4}$ $M_{\odot}$.  In Section 5, I found that the accreted mass is (1.5$\pm$0.5)$\times$10$^{-4}$ $M_{\odot}$.  In all cases, $M_{\rm ejecta}$$>$$M_{\rm accreted}$.  This answers the original prime question.  With this, the V1405 Cas WD is {\it losing} mass over each full eruption cycle.  The WD is decreasing in mass over time, so $M_{\rm WD}$ is {\it not} approaching the Chandrasekhar limit.  As such, V1405 Cas is {\it  not} a Type Ia supernova progenitor.

There are two more paths to arrive at the same conclusion that $M_{\rm ejecta}$$>$$M_{\rm accreted}$, with all three paths being independent.  The second path is easy and reliable, based on the fact that V1405 Cas is a neon nova and hence must have the eruption dredging-up the neon-rich material from the now-exposed core of the original ONe WD.  So the ejected material contains both the accreted layer plus the dredged-up mass, so $M_{\rm ejecta}$$>$$M_{\rm accreted}$.  The third path is also easy and reliable, based on the original WD being an ONe WD, necessarily starting out with $>$1.2 $M_{\odot}$, yet now the current mass is 0.60$\pm$0.10 $M_{\odot}$.  This means that over the evolution of this CV, the WD has been losing mass, and the only way to lose such mass is from many nova eruptions eroding the mass.  That is $M_{\rm ejecta}$$>$$M_{\rm accreted}$.  These two paths demonstrate that $M_{\rm WD}$ is decreasing over evolutionary time, and the WD will not approach the Chandrasekhar limit.  So we have two more confident demonstrations that V1405 Cas cannot become a Type Ia supernova.

The second progenitor test is whether it has a CO or ONe WD.  The issue is that a normal SNIa requires that the WD have a CO composition (for reviews, see Livio 2001, Tout et al. 2001, Maeda \& Terada 2016, Patat \& Hallakoun 2018).  Another way of saying this is that it is impossible for an ONe WD to explode as a supernova, because the lack of carbon means that there is not enough nuclear energy to blow up the WD.  In general, it is hard to tell the composition of a WD, with the primary exception being if the nova ejecta displays bright neon lines late in the outburst.  For such a neon nova, the high bulk abundance of the neon is impossible to be had other than by dredging up neon from an underlying ONe WD.  The only way to get a neon nova is to have the WD be of ONe composition.  V1405 Cas has been recognized by various groups (including Munari \& Valisa 2022 and Taguchi et al. 2023) to be a neon nova.  So V1405 Cas has a WD with an ONe composition.  So V1405 Cas is {\it not} a supernova progenitor.

\section{The Big Picture}

Beginning as a graduate student in the 1980s, my received knowledge on CVs and their evolution was all based on 5 popular ideas.  These 5 broad claims have all been largely adopted (with varying levels of acceptance) as the cornerstones of CV evolution by the entire CV community from the 1980s up until recently.  With numbering for the subsection below, these five claims are:   {\bf 1~~}The long-term evolution of CVs is controlled by the Magnetic Braking Model (MBM) with its schematic picture of angular momentum loss driving the changes in $P$.   {\bf 2.~~}The middle-term evolution of CVs is controlled by the Hibernation Model, where individual CVs cycle around from nova to dwarf nova to hibernation and back to nova, as driven by the $\Delta P$ of each nova eruption.   {\bf 3.~~}The WD mass of CVs is increasing over time, because $M_{\rm ejecta}$$<$$M_{\rm accreted}$.   {\bf 4.~~}CVs are the progenitors of Type Ia supernovae, as required from the fact that their WD masses will increase to reach the Chandrasekhar limit.   {\bf 5.~~}The general evolution for all CVs always changes the binaries from some relatively long orbital period (at the time the companion star first comes in contact with its Roche lobe) to a substantially shorter period as the CV ages towards its death.

All 5 claims are centered on the long-term behavior of the CV binaries.  These claims all have their most fundamental predictions and tests based on measured values for $\Delta P$ and $\dot{P}$.  Until recently, other astronomers have not even tried to measure $\Delta P$ and $\dot{P}$ for testing of these most fundamental claims on CV evolution.  With this, all direct tests of the most fundamental predictions of these claims have been ignored.  Recently, we have finally gotten 15 measures of $\Delta P$, 52 measures of long-term $\dot{P}$ for CVs, plus 25 measures of long-term $\dot{P}$ for XRBs.  Finally, we can start testing the most direct and fundamental predictions.

V1405 Cas can by itself serve as a test for all 5 claims:

\subsection{Magnetic Braking Model}

The MBM is simply a set of assumptions about a presumed power law for the AML of the binary in quiescence, which along with ordinary physics of the binary prescribes an exacting long-term evolution of the orbital period (Rappaport, Verbunt, \& Joss 1983, Patterson 1984, Knigge et a. 2011).  The label given to the AML mechanism is `magnetic braking', but this specific effect has not been seen in binaries and its critical elements (stellar wind and magnetic field from the companion) have not been observed at any significant level in any binary.  And there is no physics past the schematic level.  So the MBM is really just an empirical fit with an arbitrary power law of unknown origin (where both the slope and the constant are free parameters with huge possible ranges) such that secondary indicators for CVs are the best fit possible.  The evidence and justification for MBM are near zero\footnote{Historically, workers perceived that the MBM made successful predictions for the existence and edges of the Period Gap, and the minimum $P$ for CVs.  But the ``Standard Model'' actually gets the minimum period greatly wrong, so Knigge et al. (2011) had to increase the strength of the General Relativity effect by a factor of 2.47$\times$ so as to get agreement by the ``Best-Fit Revised Model'', so this was not a successful prediction.  Similarly for the edges of the Period Gap, Knigge et al. (2011) had to adjust parameters from the ``Standard Model'' so as to get a match with the ``Best-Fit Revised Model'', so again, adjusting the free parameters to get a fit does not count as a successful prediction.  Indeed, MBM requires that the edges of the Period Gap remain constant for all CVs, yet the reality is that the edges change greatly depending on the CV class (for example, the Gap for novae has little overlap for the Gap for Sloan CVs), and the polars do not even have a Period Gap (Schaefer 2024).  As for the existence of a Period Gap, most any model with a decreasing $P$ in evolution will produce a Period Gap.  So the MBM actually does not have any successful predictions.}, but it is the only game in town.  With no alternative in sight, the MBM was broadly adopted back in the 1980s, and by repetition has become a venerable model that is even now used as the cornerstone for all binary evolution.  What is needed is the first test of the most fundamental prediction and requirement of the MBM, that is testing whether the $\dot{P}$ is the required function of $P$.

V1405 Cas has measured $\dot{P}_{\rm pre}$=($+$1.9$\pm$1.1)$\times$10$^{-9}$ and $\dot{P}_{\rm post}$=($-$1.2$\pm$1.7)$\times$10$^{-9}$ in dimensionless units (the equivalent of seconds/second).  The pre-eruption $\dot{P}$ is {\it positive}, while the post-eruption $\dot{P}$ is some small value close to zero.  This should be compared to the prediction of $\dot{P}$ from Knigge et al. (2011).  Importantly, the MBM {\it requires}\footnote{Knigge et al. (2011) states emphatically ``Theoretically, all CVs with initially unevolved donors are expected to quickly join onto a unique evolution track, whose properties are determined solely by the mechanism for AML from the system (Paczy\'{n}ski \& Sienkiewicz 1983; Ritter \& Kolb 1992; Kolb 1993; Stehle et al. 1996). Empirically, a unique track is also necessary in order to explain the existence of a period gap with sharp edges and a well-defined minimum period.'', and ``As noted by Stehle et al. (1996), this also nicely explains the rapid convergence of CV evolution tracks characterized by different initial conditions, which is always observed in numerical studies (e.g., Paczynski \& Sienkiewicz 1983; Kolb \& Ritter 1992; Kolb 1993).''} that all $P$=0.1884 day CVs must have $\dot{P}$ exactly equal to $-$1.51$\times$10$^{-12}$.  The MBM prediction is 1.7-sigma off the pre-eruption measure and 0.7-sigma off the post-eruption measure.  The fractional uncertainties in the observed $\dot{P}$ are large.  (The large error bars are due to the relatively short stretch of data in the $O-C$ curve, which is due to the new plan for measuring $\Delta P$ with its small interval of pre-eruption archival data.)  Nevertheless, the pre-eruption $\dot{P}$ is positive, and a positive $\dot{P}$ is impossible within MBM.

For the bigger picture, the MBM has already been refuted by the collections of $\dot{P}$ measures for 52 CVs of all classes (Schaefer 2024) and 25 XRBs of all classes (Schaefer 2025b).  That is, nearly all CVs and XRBs have observed $\dot{P}$ that are 1-to-3 orders-of-magnitude in error.  And roughly half of the $\dot{P}$ values are {\it positive}, with such being impossible in the MBM (e.g., Knigge et al. 2011).  These are the first tests of the primary prediction and requirement of the MBM, and the model has always failed badly.  The MBM is dead.

\subsection{Hibernation model}

The Hibernation model controls the middle-term evolution of all CVs, with the hibernation effect driven by a very large positive $\Delta P$ for each nova eruption (Shara et al. 1986).  The model predicts that each individual CV cycles around from being an ordinary classical novae, to a dwarf nova state, to a state of detached hibernation, then back to a dwarf nova state, and reaching the nova state, to start the cycle all over again.  This is an alluring model, but a significant fraction of the CV community has not been convinced.  It is hard to test the model, because the changes are on the century or longer timescale.  Some novae do show a slow fading (e.g., V603 Aql since 1918, Johnson, Schaefer, et al. 2014), a number of others show century-long steady {\it brightening} (e.g., Q Cyg since 1876), while most do not show any systematic long-term fading even out past a century, and the whole V1500 Cyg class of novae have the quiescent system long after the nova being substantially {\it brighter} than the pre-eruption quiescence (Schaefer \& Collazzi 2010).  Hibernation predicts that Z Cam dwarf novae might show old nova shells far away from the star, and this prediction has received four impressive confirmations ( Shara et al. 2007, 2017a, 2017b, 2024, Miszalski et al. 2016).  Unfortunately, the existence of extended nova shells around dwarf novae is also predicted for the case with no hibernation (all CVs have novae sooner or later), so the fulfillment of the prediction is actually ambiguous.  So until recently, the observational evidence has not been decisive.

The critical prediction of Hibernation is that the novae have large-and-positive $\Delta P$, with this being required to induce the state of hibernation.  That is, if $\Delta P$$>$0 then the binary must separate and the accretion rate will decline to make the system dimmer, and if $\Delta P$$\gg$0 then the induced binary separation might be large enough so that the dimming will be large enough to qualify the star as having gone into a hibernation state as a detached binary.  Schaefer (2020a, 2020b, 2023b) calculates that Hibernation requires that the observed $\Delta P$/$P$ be larger than $+$1000 for short period systems (like V1405 Cas) and larger than $+$10,000 for long period systems.  So for the case of V1405 Cas, with $\Delta P$/$P$ of $+$66, there is no chance for a separation adequate to make the system fade to the point where we could call it `hibernation'.  Indeed, for the observed value, the binary separation increase is so small that there can be no measurable fading of the quiescent nova from before-to-long-after the nova.  So V1405 Cas provides a refutation of the Hibernation model, at least for one ordinary nova.

It turns out that the Hibernation model is already long gone.  Of the 12 nova with measured $\Delta P$, 5 of them are {\it negative}.  This is `anti-hibernation'.  Of the 7 systems with positive-$\Delta P$, they all are from $>$18$\times$ to $>$900$\times$ too small for hibernation, so that the observed $\Delta P$ does not even allow any measurable dimming.  This is an utter failure of the primary requirement and prediction of the Hibernation model.

\subsection{$M_{\rm WD}$ is {\it increasing} over time}

Many workers in the CV community have the general idea that the WD mass is increasing in time.  Simplistically, CVs all have the WD being piled higher and deeper with accreted gas, and it is easy to overlook the effects of nova eruptions.  This is all wrapped up in the SNIa progenitor problem, where Single-Degenerate (SD) model enthusiasts need CVs to have $M_{\rm WD}$ growing to the Chandrasekhar limit.  Nova theorists have become polarized.  So we see entrenched groups claiming that CVs have growing WD masses (e.g., Starrfield et al. 2025, Hillman et al. 2016) versus other groups that calculate that CV WD masses are not growing (e.g., Yaron et al. 2005, Idan, Shaviv, \& Shaviv 2013).  These indecisive arguments between modelers on the many astrophysical and technical issues (e.g., Kato, Hachisu, Saio 2017) are irrelevant, because we now know that all models have completely missed the dominant mass ejection mechanism.  And observers cannot effectively address the claim, because all the traditional methods have $>$100$\times$ real uncertainties.

For the case of V1405 Cas, we have a clear result.  We have $M_{\rm ejecta}$$>$$M_{\rm accreted}$, so $M_{\rm WD}$ is decreasing over each eruption cycle.  Further, V1405 Cyg is a neon nova, so the nova eruption is dredging up material from the mantle of the underlying WD, which means that the eruption is ejecting much more than the layer of recently accreted gas.  Indeed, the original ONe WD must have $>$1.2 $M_{\odot}$ (all ONe WDs form above this limit) while the current ONe WD is at 0.60 $M_{\odot}$, so the V1405 Cas WD has lost over half of its original mass.  With these three proofs, we know that V1405 Cas violates the general claim that CVs have their WD masses increasing over evolutionary timescales.

For the broader case, it turns out that {\it all} CVs with real evidence have the WD {\it decreasing} in mass.  I know of four broad and general lines of evidence that all CV WDs are losing mass over time:  

{\bf First,~}for the 6 novae with unambiguous (i.e., positive) $\Delta P$ measures, all 6 have $M_{\rm ejecta}$$>$$M_{\rm accreted}$.  In the case of V1405 Cas, the ejecta is $>$2$\times$ larger than the break-even level.  The SD prototype systems are 26$\times$ over for U Sco (Schaefer \& Myers 2025), $\gg$5$\times$ over for V445 Pup (Schaefer 2025a), 540$\times$ over for T CrB (Schaefer 2025c), and $\gg$11.3$\times$ over for T Pyx.  Ordinary nova BT Mon has the ejected mass $\gg$0.8$\times$ over that accreted.  This is telling us that the eruption physics of a wide variety of systems always ejects much more mass than is accreted, and this physics conclusion is then applicable to all CVs.    

{\bf Second,~}roughly one-third of all novae are neon novae\footnote{This one-third figure is taken from Truran \& Livio (1986).  As an update, I have compiled an exhaustive list of 37 novae that have abundance analyses, of which 46\% are neon nova.}, and these necessarily are ejecting much more mass than is being accreted.  To access the only source of bulk neon (i.e., the core of an ONe WD), the original crust and mantle of the ONe WD must already have eroded away from prior nova explosions.  To emphasize the requirement of mass loss, Schaefer (2025d) finds that 76\% of all neon novae have $M_{\rm WD}$$<$1.2 $M_{\odot}$, with this limit being the lower limit for the original ONe WD mass, so we know that a large fraction of novae in the sky are necessarily having their WD masses being whittled down over time.  

{\bf Third,~}with my collection of all published formal abundance analyses for nova ejecta, the result is that {\it all} 34-out-of-34 novae have supersolar abundances of He, C, N, O Ne, Mg, and/or Al.  These super-solar abundances cannot come from the companion star or from nuclear burning of the accreted material (Truran \& Livio 1986), so this ejecta material can only have a source from the underlying WD (De Ger\'{o}nimo et al. 2019).  For a general example, a supersolar abundance of carbon and oxygen can only come from the core of a CO WD or from the mantle of an ONe WD.  For a particular example, V838 Her has 37.9$\times$ solar abundance of nitrogen, 52.5$\times$ solar of neon, and 29$\times$ solar of aluminum, and such can only have a source from the core of an ONe nova, where the outer crust and mantle have already been stripped away from nova eruptions.  To get this heavy-element enriched material out into the ejecta, the only mechanism is for the nova eruption to dredge-up and eject the underlying WD material.  Therefore, the nova has recently ejecting dredged-up mass in addition to the accreted layer (so $M_{\rm ejecta}$$>$$M_{\rm accreted}$).  Further, many prior eruptions have eroded the WD surface down to its interior, again requiring the WD mass to be decreasing.  So the fact that all tested novae have supersolar abundances of heavy-elements demonstrates that all 34-out-of-34 novae are ejecting more mass than accreted over each eruption cycle.  

{\bf Fourth,~}in my compilation of WD masses for 305 CVs with $P$$<$0.60 days for main sequence companions\footnote{This is ignoring the 24 CVs with sub-giant companions and the 6 CVs with red giant companions, because they certainly have different evolutionary paths from all the usual CVs with low-mass companions.}, I see that only {\it two} nova\footnote{The exceptions are T Pyx and V838 Her, with both being proven to be young CVs, with little time to erode a WD to low mass.} have masses $>$1.28 $M_{\odot}$.  The WD mass distribution has a sharp dropoff from 1.20--1.28 $M_{\odot}$, and the drop is by a factor of $\sim$16$\times$ and is highly significant\footnote{For CVs with main sequence companions, 31 have masses from 1.00--1.09 $M_{\odot}$, 33 have masses from 1.10--1.19 $M_{\odot}$, 18 have masses from 1.20--1.29 $M_{\odot}$, and only 2 have a mass between 1.30--1.39 $M_{\odot}$.}.  I know of no selection effect or WD formation bias that can create this sharp cutoff.  If CV WD masses are {\it rising} with evolution, then all the many CV binaries that come into contact must track into the void on their way to reaching the Chandrasekhar mass.  So the increasing-$M_{\rm WD}$ hypothesis predicts many CVs with masses $>$1.28 $M_{\odot}$, which is refuted by the data.  However, if CV WD masses are {\it decreasing} with evolution, then all the ONe WDs with original masses $>$1.28 must speedily vacate the void, evolving into the many neon nova with WD masses below 1.2 $M_{\odot}$, while there are no CVs with WD masses above the Chandrasekhar limit that can evolve to fill the void.  This is a proof that CV WDs in general are losing mass over time.

\subsection{CVs are Type Ia supernova progenitors}

The `Single-Degenerate' (SD) model for SNIa was the original idea going back to the 1960s, and it was the only model up until 1984 when the `Double-Degenerate' (DD) model came out (Webbink 1984, Iben \& Tuttkov 1984).  From the earliest days even to now, the SD solution seems inevitable, because CVs are the full realization of progenitors featuring WDs that are being fattened-up to the Chandrasekhar mass.  The SD versus DD discussions have been `enthusiastic' and have polarized astronomers.  The broad importance of this issue has turned the SNIa progenitor problem into one of the grand challenges for astrophysics for the last four decades (Livio 2000, Maoz, Mannucci, \& Nelemans 2014, Ruiter \& Seitenzahl 2025).  All of the SD models identify the CVs as the progenitors, and they take many classes of CVs as their progenitor prototypes, including U Sco and T Pyx for recurrent novae, T CrB for symbiotic stars, and V445 Pup for helium novae.

The case of V1405 Cas can be used to test the general idea that CVs will evolve into SNIa.  Section 6 provides a conclusion of high certainty -- V1405 Cas will not become a SNIa.  The proofs are that $M_{\rm ejecta}$$>$$M_{\rm accreted}$ as measured from $\Delta P$, that the 2021 eruption was a neon nova and hence the underlying WD must be of ONe composition, and the current WD (with mass 0.60 $M_{\odot}$) must have been losing mass (down from its original $>$1.2 $M_{\odot}$) over its whole lifetime as a CV.

More broadly, the possibility of CV progenitors is already gone:  {\bf First},~from Section 7.3, all 6-out-of-6 novae with $\Delta P$, all 39-out-of-39 neon novae, and all 34-out-of-34 novae with heavy element enhancements are already proven to have the WD losing mass over evolutionary time, and thus cannot be progenitors.  These proven non-progenitors span all classes of CVs and span the entire range of CV properties.  This is a demonstration that the physics of nova eruptions is always producing explosions that dredges up and ejects more material than is in the accreted layer.  The physics of eruptions does not change over small regions of parameter space, so the untested CVs will surely have the same dredge-up results as does a `neighboring' CV with nearly identical conditions that is already proven as a non-progenitor.  So all the untested CVs are non-progenitors.

{\bf Second,~}all CV progenitors must have a companion star at the time they explode as a SNIa, perhaps a red giant, a subgiant, or a main sequence star.  These companions will survive the nearby explosion, with a luminosity largely similar to before the explosion.  So a deep enough search before/during/after the supernova must always detect the companion.  The companions can be detected before the eruption by looking in deep {\it HST} images of the site.  During the eruption, the companion is recognized from the ablated and entrained hydrogen and helium as bright nebular emission lines, as well as by the  radio and X-ray luminosity from the ejecta shocking against the companion's stellar wind, and by the  so-called `Kasen peak' that is visible from one hour to 2 days after the explosion as a peak in the light curve that is comparable to the regular peak.  After the eruption, the battered companion will be visible at the center of young remnants.  With a total of 689 normal SNIa events searched deeply, zero were found that could possibly have any red giant companion, zero that could possibly have a subgiant companion, and zero were found that could possibly have a main sequence star more massive than our Sun.  So to the level of $<$0.16\%, SNIa are not produced from CVs with red giant stars, subgiant stars, or massive main sequence stars.

{\bf Third,~}the CV stellar masses that can possibly produce a SNIa are strongly constrained.  For ordinary CVs, we have that $M_{\rm WD}$+$M_{\rm comp}$ must be greater than the Chandrasekhar mass for there to be any possibility of a supernova.  For CVs with companions that have a helium atmosphere, the possibility of sub-Chandrasekhar explosions makes the requirement that $M_{\rm WD}$+$M_{\rm comp}$$>$0.9 $M_{\odot}$ for any viable progenitor.  For the many CVs with measured $M_{\rm WD}$$<$0.5 $M_{\odot}$, the WD must either have started out $<$0.5 $M_{\odot}$ in which case it is a helium WD that cannot explode or have started out $>$0.5 $M_{\odot}$ in which case the CO WD is decreasing in mass from evolution and will never explode.  These three requirements eliminate nearly 90\% of known CVs, including the AM CVn stars, the symbiotic stars, and nearly all of the ordinary CVs.

{\bf Fourth,~}with the above proofs, nearly all CVs are demonstrated to be impossible progenitors.  The few remaining CV (those not already tested) are in a small region of parameter space, centered around  the famous U Gem itself.  So, I have not eliminated all individual CVs as proven non-progenitors, but I am close.  This result can be used with Occam's Razor to generalize to {\it all} CVs.  A CV-progenitor enthusiast should test and compare two hypotheses.  The first hypothesis is that all CVs are the same class of non-progenitors.  The second hypothesis is that some small and unrecognized class of CVs  does actually make for SNIa up in the sky.  This second class must be new and rare, and its properties must fit into a rather tight set of requirements (c.f., the first three points above).  By Occam's Razor, the second hypothesis is strongly denied.  So, CVs are not SNIa progenitors.  And V1405 Cas provides good evidence in support of this broad conclusion.   
 
\subsection{CV evolution runs from long-$P$ to short-$P$}

The longest and strongest idea about CV evolution is that $P$ changes from some relatively long-period at the time the binary comes in contact down to some relatively short-period at the end of the star.  This perceived universal property is from the earliest days, and I first had it schooled into me during professorial discussions and seminars at my university and next door, with the definitive papers such as Rappaport, Verbunt, \& Joss (1983), Patterson (1984), Knigge, Patterson, \& Barraffe (2011) drumming in the relentless shortening in the period.  I have never heard the basic idea questioned anywhere by anyone.

The measured evolution of $P$ for V1405 Cas can be used to test this general expectation.  For the best fitting broken parabola, $P$ was 0.1883857 days in 2013 and 0.1884039 days in late 2024.  So with the longest baseline available, V1405 Cas has {\it increased} its orbital period.  This is contrary to the oldest and strongest idea about CV evolution.  Now, for any one case, the lengthening of the period can be reasonably shrugged off as a deceptive transient not representative of the run over the longest timescales.  But this sample of one has made me realize that the universal long-to-short evolution paradigm can and must be tested.

The original idea was that the binary AML mechanism is inevitably grinding down the orbital angular momentum, driving the orbit to shorter and shorter periods.  This AML applies both to CVs between eruption (with this mechanism being unknown, and certainly not the MBM) and to CVs with their inevitable nova events (say, with the frictional AML on the companion as it plows through the ejecta and envelope around the WD).  These circumstances of AML make for the general idea that CV periods must be grinding down.

With new realizations for period-change mechanisms, the $P$-shortening paradigm should be revisited.  I can think of three strong effects for increasing-$P$ that might be dominant:  {\bf First}, the ordinary mass transfer from the companion star to the higher-mass WD must necessarily {\it increase} the $P$.  This is a large effect over the lifetime of CVs.  If half of $M_{\rm comp}$ is transferred to the WD, then the period will roughly {\it double}, with the exact size of the effect depending on the initial mass ratio.  {\bf Second}, the mass ejected from the WD during a nova eruption will necessarily {\it increase} the $P$ (see Equation 5).  This is a large effect over the many eruptions in all CV lifetimes.  For cases like V1405 Cas where the WD has ejected nearly half of $M_{\rm WD}$, the period will roughly {\it double}, with the exact size of the effect depending on how much mass is ejected.  So these two effects will roughly quadruple the $P$ over the CV lifetimes.  This quadrupled-$P$ is in competition with the shortening of the period from AML effects.  {\bf Third}, the observed $\dot{P}$ for nearly half of the CVs is positive, and often large (Schaefer 2024).  For at least these systems, the period is increasing fast.  We do not understand the mechanism for the positive-$\dot{P}$, but the orbital periods are increasing anyway.  So empirically, we know that some dominant mechanism is running opposite to the claim of universal period-shortening.  Current CV evolution models are still operating with the disproven MBM, so none are including any sort of a realistic AML mechanism, so all current programs and papers and results are missing the mechanism that makes $P$-increase dominant for nearly half the CVs.

I think that current CV evolution programs do not account for any of these three effects.  I do not know whether the mass-transfer or AML effects will dominate.  But now that the question is raised, the shortening-$P$ evolution no longer seems inevitable.  Rather, perhaps most CVs start out with some $P$ and evolve to a {\it longer}-$P$.  Or perhaps some CVs have lengthening-periods while others have shortening-periods.

Recently, our community has a large sample of century-long $O-C$ curves, and these can be used to test whether the periods are indeed shortening.  The best and only sample is the 52 CVs in Schaefer (2024).  Out of the systems with main sequence companions and no nova eruption in the observed interval, 18 have the period increasing from beginning to end, while 23 had the period decreasing from beginning to end.  Out of the 12 nova systems with measured $\Delta P$, 7 have period increases, while 5 have period decreases.  So for CVs in general, the number of increasing-$P$ systems is comparable to the number of decreasing-$P$ systems.  This large sample provides an empirical disproof of the universality of the shortening-$P$ claim.  With 41 CVs of all types, we cannot attribute the observational result to selection effects or small number statistics or special cases.  So the universal shortening-$P$ paradigm appears to be lost.

\section{Conclusions}

V1405 Cas is an ordinary classical nova discovered by Y. Nakamura, peaking at $V$=5.2, with 9 large-amplitude jitters up-and-down over a long peak.  The decline rate is $t_3$=175 days, for a light curve class of J(175).  This is amongst the slowest known novae, pointing to a very low mass WD, taken to be 0.60$\pm$0.10 $M_{\odot}$.  The eruption spectra showed V1405 Cas to be a hybrid nova, primarily with a spectral classification of Fe II, although with HeN appearances briefly at the start at end of the eruption.  Neon lines appear prominently, making this a neon nova.  The quiescent pre-eruption nova was actually discovered in 2020 by Z. Henzl, as a variable star with a sinewave modulation.  Post-eruption {\it TESS} light curves demonstrated a period of 0.1884 days, dominated by irradiation effects on the 0.43$\pm$0.04 $M_{\odot}$ companion star.  I have pulled out 20 accurate times of photometric minima from light curves from {\it TESS}, AAVSO, APASS, ASAS-SN, ATLAS, and {\it Gaia}, stretching from 2013--2025.  From these I derive the fractional period change across the 2021 eruption to be $\Delta P$/$P$ of $+$66 parts-per-million (with the one-sigma range of 48--87 ppm).  The steady period changes are $\dot{P}_{\rm pre}$=($+$1.9$\pm$1.1)$\times$10$^{-9}$ and $\dot{P}_{\rm post}$=($-$1.2$\pm$1.7)$\times$10$^{-9}$.

With the observed $\Delta P$/$P$ of $+$66 ppm, I can calculate the mass ejected by the nova in 2021.  But first, I have to estimate the period change caused by the frictional angular momentum loss by the binary during its eruption, $\Delta P_{\rm aml}$, as the companion orbits inside the hot envelope thrown up around the WD.  Current theory is not able to model $\Delta P_{\rm aml}$, so instead I have derived $\Delta P_{\rm aml}$ for two twin novae (HR Del and RR Pic) which are closely similar to V1405 Cas.  With this, I find that $\Delta P_{\rm aml}$/$P$ for V1405 Cas is roughly $-1400$ ppm, but with an extreme range of $-$490 to $-$8700 ppm.  Then, with $\Delta P_{\rm ejecta}$=$\Delta P$-$\Delta P_{\rm aml}$ and $\Delta P_{\rm ejecta}$=2$P$$M_{\rm ejecta}$/($M_{\rm WD}$+$M_{\rm comp}$), the best estimate is that V1405 Cas ejected 7.5$\times$10$^{-4}$ $M_{\odot}$, with an extreme limit of $M_{\rm ejecta}$ $>$ 2.9$\times$10$^{-4}$ $M_{\odot}$.  From the nova trigger condition, the mass of the gas accreted before the 2021 eruption was (1.6$\pm$0.4)$\times$10$^{-4}$ $M_{\odot}$.  With this, $M_{\rm ejecta}$$>$$M_{\rm accreted}$, the WD is losing mass over each eruption cycle, and V1405 Cas has no path to become a Type Ia supernova.

V1405 Cas can be used to test 5 separate claims that have dominated the entire field of CV evolution since the 1980s:   {\bf 1.~}For V1405 Cas, the measured pre-eruption $\dot{P}$ value is positive, which is impossible within the MBM, although this is not strong due to relatively large error bars.  But the MBM is already dead because the large database of $\dot{P}$ measures for 77 CVs and XRBs has already refuted the old venerable model.  {\bf 2.~}V1405 Cas has the measured $\Delta P$ that is 20$\times$ smaller than required by the Hibernation Model, so that any dimming effect cause by the increased binary separation is lost in the usual flickering.  But the Hibernation Model is already dead because the many measures of $\Delta P$ all show that the kicks to the orbit are in the wrong direction about half the time, and the rest of the systems have kicks that are 1--3 orders-of-magnitude too small to create a hibernation state.  {\bf 3.~}V1405 Cas provides a counterexample to the claim that CV WD masses are generally increasing over evolutionary time, both because this nova has $M_{\rm ejecta}$$>$$M_{\rm accreted}$ (from the $\Delta P$) and was a neon nova.  This refutation is just part of a much larger picture where all measures show $M_{\rm ejecta}$$>$$M_{\rm accreted}$ for most CVs,  if-not-all CVs.  {\bf 4.~}V1405 Cas is certainly not a SNIa progenitor, because $M_{\rm ejecta}$$>$$M_{\rm accreted}$ and because it was a neon nova.  But the SD models with CV progenitors are already refuted by all the prototypes having mass-losing-WDs and ONe WDs,  Further, CV progenitors with red giant companions, subgiant companions, or solar-mass main sequence star companion are refuted by the many and extremely deep searches for the companions before/during/after normal SNIa eruptions.  {\bf 5.~}V1405 Cas has significantly {\it increased} its orbital period (by $+$75 ppm) over the observing interval of 2013--2025, with this being contrary to the pervasive claim that ordinary CV periods universally evolve from long-$P$ to short-$P$.  This result for V1405 Cas is just a part of the bigger picture, for 41 CVs with long $O-C$ curves, about half of them are {\it increasing} their $P$ from start to end (Schaefer 2024), so the universality of the declining-$P$ claim is questionable.

V1405 Cas is just an ordinary nova, but it has high importance now with the measure of $\Delta P$.  With this period-change information, V1405 Cas contributes to the refutations of the 5 claims that have dominated models of CV evolution since the 1980s.  The failures of these claims have only become apparent recently, because previously our community did not have any effective data set of $\Delta P$ and $\dot{P}$ measures to serve as the only direct tests of CV evolution.

These 5 failures impeach the critical aspects of all prior papers on CV and XRB evolution, demographics, and population synthesis.  Our community desperately needs realistic models for the ordinary angular momentum loss during quiescence, the angular momentum loss during nova eruptions, and the mass ejection during nova eruptions.

\begin{acknowledgments}

Paul Barrett (George Washington University) was helpful for obtaining the ATLAS data, and for learning about the DNO and QPOs.
  
\end{acknowledgments}

\vspace{5mm}
\facilities{AAVSO, TESS, Gaia, ASAS-SN}


{}

\end{document}